# Adhesion Induced Instabilities and Pattern Formation in Thin Films of Elastomers and Gels

Manoj K. Chaudhury[1,*], Aditi Chakrabarti[1] and Animangsu Ghatak[2]

[1]Department of Chemical and Biomolecular Engineering, Lehigh University, Bethlehem, PA 18015, United States.

[2]Department of Chemical Engineering, Indian Institute of Technology, Kanpur, UP 208016, India.

**Abstract.** A hydrostatically stressed soft elastic film circumvents the imposed constraint by undergoing a morphological instability, the wavelength of which is dictated by the minimization of the surface and the elastic strain energies of the film. While for a single film, the wavelength is entirely dependent on its thickness, a co-operative energy minimization dictates that the wavelength depends on both the elastic moduli and thicknesses of two contacting films. The wavelength can also depend on the material properties of a film if its surface tension has a pronounced effect in comparison to its elasticity. When such a confined film is subjected to a continually increasing normal displacement, the morphological patterns evolve into cracks, which, in turn, govern the adhesive fracture behavior of the interface. While, in general, the thickness provides the relevant length scale underlying the well-known Griffith-Kendall criterion of debonding of a rigid disc from a confined film, it is modified non-trivially by the elasto-capillary number for an ultra-soft film. Depending upon the degree of confinement and the spatial distribution of external stress, various analogs of the canonical instability patterns in liquid systems can also be reproduced with thin confined elastic films.

Correspondence should be addressed at: mkc4@lehigh.edu

## 1. Introduction.

Instability driven pattern formation is abundant in nature. A gently flowing liquid thread becomes unstable to form droplets [1]. Solidification of microscopic super-cooled water drops lead to the formation of dendrites [2,3]; those become the delightful snowflakes. Differential velocities of two fluid streams often lead to exquisite wavy patterns [4] that are occasionally observed with wind blowing over the surface of a sea or even over a cloudy sky. Fingering patterns [5], often accompanied by highly branched structures [6], develop on the surface of a high viscosity liquid while being displaced by a fluid of lower viscosity. These and many other instability driven patterns are so common in nature that an eminent physicist, R. Sekerka, expressed his optimism [7] as: "If there isn't already an instability, you figure God will provide one".

While these broad classes of morphological patterns are consequences of the imbalance of the thermal, chemical and momentum fluxes across a fluid interface, there are also numerous examples of instabilities observed with elastic systems, where a flow is not necessarily present. Classical examples include Euler buckling instability in a rod and wrinkling instability in a thin film in the presence of an axial load or a lateral compressive force [8-12]. A sufficiently compressed soft gel or an elastomer leads to folding and/or creasing patterns on its surface, which have attracted a great deal of attention [13-19] of not only the material scientists but also the biophysicists as they offer interesting avenues to study the complex phenomenology [20-23] of growth and forms in biologically evolving systems. There also exists a great amount of work focusing on the instabilities engendered [24-26] in pre-stressed epitaxial metal or semiconductor films as a consequence of the destabilizing surface diffusion and stabilizing surface tension forces. For example, a thin plastic film bonded to an elastic substrate shows surface undulations [27], which depend upon the mismatch of internal strains.

Various types of morphological [28-53] evolutions have also been observed in such soft materials as foams [28], elastomeric [29-43] and viscoelastic polymers [44-53] under confinement, when a suitable destabilizing force (e.g. a mechanical, electrical or van der Waals force) acts upon it. The morphological features of these patterns depend on their geometric and, and in some cases, on their material properties such as elastic modulus and surface tension, while the ratio of the relaxation to the observational time dictates whether the instability to be observed is related to an elastic or a viscous deformation. In these contexts, the role of an important length scale – the elastocapillary length (ratio of surface tension to elastic modulus) was identified that competes or co-operates with the geometric scales in the wavelength selection process. The importance of this length scale has become even more evident in several recently discovered new phenomena [54-68], which are blossoming into an exciting field of research in soft matter physics.

The subject related to elastic instability in compliant matters is rather vast, and it will be too daunting to try to review these developments in a single article. We narrow down the focus of this article onto a particular type of instability that is the adhesion induced pattern formation in constrained soft films, which was discovered independently at Lehigh University [29] and the University of Ulm [30]. The ensuing morphologies in soft films are not only delightful, but studying their properties is essential in developing fundamental insights into the nucleation of crack, fracture and friction at soft material interfaces. The theoretical understanding of the adhesion induced instability arising from the competition between elasticity and the long range van der Waals force (during the bonding phase) was largely contributed by the groups originating from the University of Ulm [30] and IIT-Kanpur [31]. The analysis required for the de-bonding state, which is the subject of this article is, formally, somewhat different from the previous studies as will be discussed below.

To motivate our discussion, let us consider a thin soft film of a rubber or a gel strongly bonded to a solid support. If one brings a flat rigid disc into contact with such a thin film and attempts to either pull it apart or slide it against the film, the interface degenerates into a worm-like pattern or gets populated with numerous bubbles. In the shear mode [39], the bubbles travel across the area of contact, often with a speed that is almost thousands times larger than that of the slider itself ! These are the surrogates of dislocations that appear in crystalline solids, which facilitate relative sliding of the rigid disc against a soft substrate. With the application of a force normal to the interface, the worm like patterns grow, thereby increasing the compliance of the system that allows easy removal of the rigid disc from the soft film. In a variation of this experiment, bubbles can be trapped at the interface by gently dropping a thin plate parallel to a soft film [37].

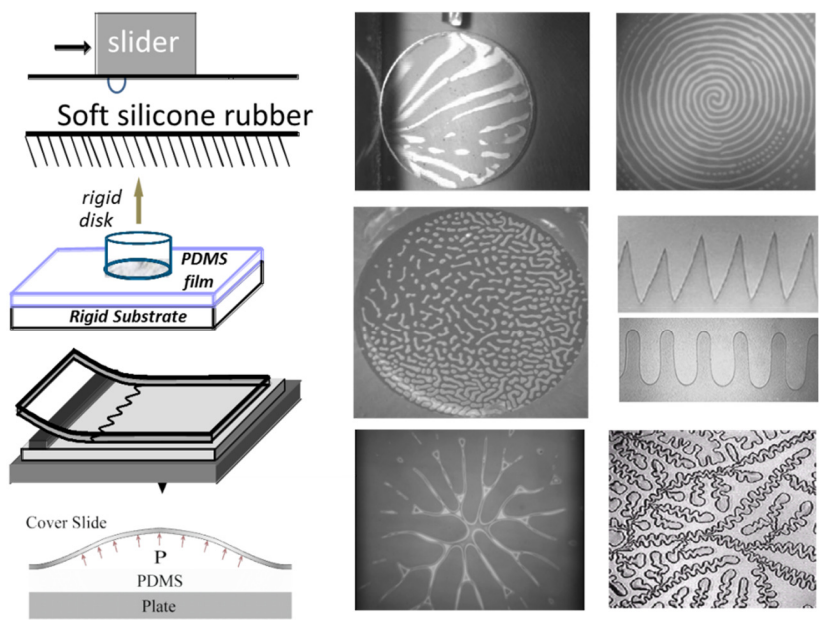

**Figure 1**. A repertoire of adhesion induced patterns that form at the interface of a thin film and a contactor in such experiments as de-bonding, sliding, peeling and the healing of a blister. Details of each of these patterns are discussed in the text.

These blisters evolve into self-generated hydraulic channels that control the transport of the trapped air.

With slight modifications of the experimental conditions, other types of interesting patterns can also be produced. For example, when a flexible cantilever is peeled from a thin soft film [29], U-shaped fingering patterns are formed that reminds us of the generic Saffman Taylor instability; but, these patterns in elastic films neither grow nor decay with time. When a rigid disk is rotated about its axis while keeping it in contact with the thin elastic film [32], the worm like instabilities or isolated cavities coalesce and evolve into such intriguing patterns as concentric rings and/or spirals. Various other types of elegant patterns can also be formed by controlling the distribution of the stress as modified by the flexibility of the plate [32], pre-existing defects and/or the mode of application of an external force.

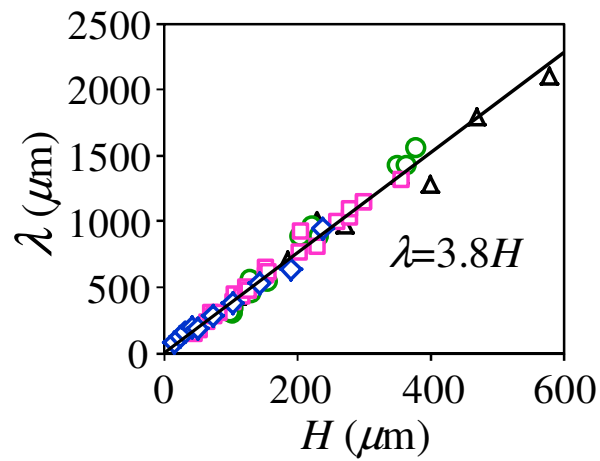

**Figure 2.** Wavelength of instability from large variety of experiments (figure 1) involving films of different shear modulus, $\mu = 0.25 - 2.0$ MPa and contactors of different rigidity all fall on a single master line. The symbol (▢) indicates peeling instability (Ref. 29, Fig. 3), (◆) indicates instabilities in a blister (Ref. 37, Fig. 1) and (○,▲) indicate cavitation instability (Ref. 38, Fig. 4a).

However, irrespective of the ways these morphologies develop in a single soft film, the unique feature with all these patterns is that each can be characterized by a dominant wavelength ($\lambda$) that is independent of all the material properties, except the film thickness ($H$). As figure 2 reveals, $\lambda$ scales with $H$ as $\lambda = 3.8H$. More complex scenarios, nevertheless, emerge with two soft films, which will be discussed later in this article.

## 2. Role of Instability in the Adhesive Fracture

The importance of the worm like instability in the fracture of an adhesive joint can be understood with the following exposition. The stress needed to debond an adhesive joint made up of two infinitely rigid materials is predicted to be: $\mathcal{A}/6\pi\ell_o^3$, $\mathcal{A}$ being the Hamaker constant of interaction and $\ell_o$, the molecular separation distance between the two surfaces. Starting with the original work of Griffith [69], we know today that fracture is a problem of thermodynamic instability, in which part or all of the stored elastic energy of deformation is used up in the creation of two free surfaces. Thus, if a particular system is not free to undergo significant deformation, a very large amount of stress would be needed to initiate and propagate a crack. For soft rubber in contact with a flat surface, this stress can be so large that it would reach or even exceed its cohesive stress. However, we know from our experience that interfacial failure does occur in such systems at a stress, which is often smaller than its elastic modulus. This is possible as multiple flaws develop co-operatively at the interface, which increases the compliance of the system, thus the stress intensity, thereby reducing the stress required to initiate interfacial fracture. This co-operative development of flaws mediated by elastic strain field leads to one particular type of instability as discussed below. In order to illustrate the theme, we first consider the case of a thin soft elastomer or a gel sandwiched between two rigid flats.

The relationship between stress ($\sigma$) and the deflection ($w$) of the surface can be described by the well-known equation of Kerr [70], which is presented [71] below (equation 1) for the case of perfect bonding between the thin film and the support, but by considering no friction between the film and the contactor to be removed from it:

$$\left[\frac{1}{2}H(1+\nu)-\sin(H\nabla)\cos(H\nabla)\left(\frac{1}{2\nabla}(1+\nu)(3-4\nu)\right)\right]\frac{(1-\nu^2)\sigma}{E}$$
$$=\left[\frac{1}{4}(1+\nu)(3-4\nu)\sin(H\nabla)\sin(H\nabla)-(1-\nu^2)(1-\nu)+\frac{1}{4}(1+\nu)H^2\nabla^2\right]w \tag{1}$$

Here $E = 2\mu(1+\nu)$ is the Young's modulus, $\nu$ is the Poisson's ratio and $\mu$ is the shear modulus respectively. $\nabla$ represents the differential operator, which is equal to d/dx in one dimension. The trigonometric functions can be expressed as the power series of $H\nabla$, and equation 1 can be solved by retaining the first three or four terms of the expansions. Let us first consider the case of a flat ended cylindrical indenter (diameter $2a$) being pulled from an incompressible elastomeric film (Appendix B, Movie 1) that undergoes a uniform vertical deflection $w_0$. In this problem, the normal stress [$\sigma(r)$, in equation 3] can be obtained from the solution of equation 2:

$$w_0 = -\frac{H^3}{E}\frac{1}{r}\frac{\partial}{\partial r}\left[r\frac{\partial\sigma}{\partial r}\right], \tag{2}$$

or, $$\sigma(r) = \frac{Ew_0}{4H^3}(a^2 - r^2) \tag{3}$$

The total energy of the system is comprised of the potential ($U_p$), elastic ($U_E$) and adhesion ($U_a$) energies, where the magnitude of $U_p$ is twice that of $U_E$, but is negative. We thus have:

$$U_T = U_P + U_E + U_a$$
or, $$U_T = -\frac{4F^2H^3}{\pi Ea^4} - \pi a^2 W_a \tag{4}$$

Where, $F$ is the applied force and $W_a$ is the work of adhesion. Using the critical crack growth condition at a fixed load, $(\partial U_T/\partial a = 0)$, we obtain the average stress needed to debond the cylinder from the thin film (similar to equation 72 of [72]) as:

$$\sigma_c = \left(\frac{a}{2H}\right)\left(\frac{W_aE}{2H}\right)^{1/2}. \tag{5}$$

However, as $a >> H$ for a thin confined film, this stress is quite large – it can easily be much larger than the modulus and even the cohesive strength of the film. Thus an adhesive separation at the cylinder - film interface by a homogenous vertical deformation is a highly unlikely mode of failure. Even if a single defect of size $R^* \sim (\Delta\gamma\mu/\sigma^2)$ could be nucleated at the expense of the creation of surface energy $\Delta\gamma$, by thermal fluctuation, the required energy barrier $\sim (\Delta\gamma)R^{*2}$ of nucleation is enhanced so much by the confinement parameter by ($a/H$) that no normal fluctuation in the system will be able to surpass it. The system would therefore opt for an alternate route to failure, if *Nature* allows that.

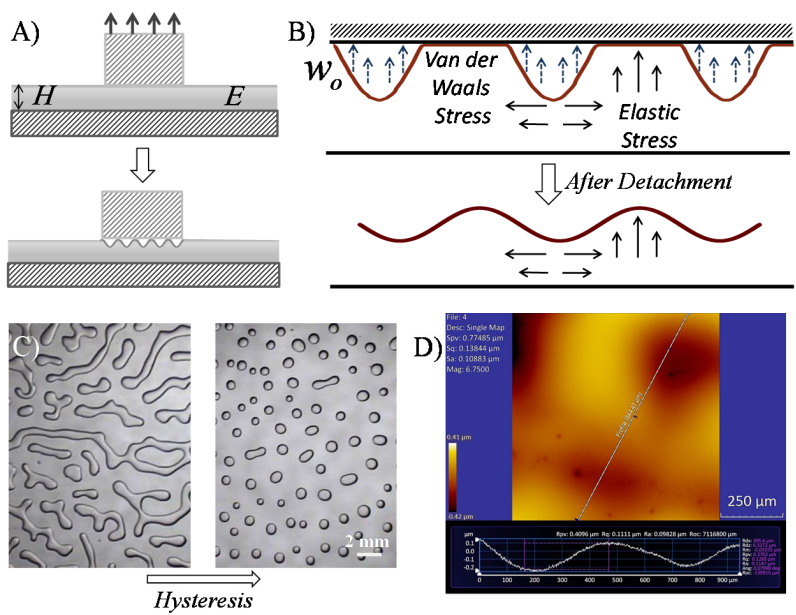

**Figure 3.** A schematic of surface undulation (B) when a rigid block is being pulled normal to the interface (A). When a glass plate is peeled from a thin ultra-low modulus hydrogel film, worm like instability develops (C, left). However, when the plate is completely peeled from the gel and re-positioned on the same hydrogel film, bubble like instability is formed (C). A profilometric image of the gel's surface soon after the plate was removed showing sinusoidal undulation.

The strategy, as selected by *Nature* herself, is the formation of co-operative defects that increases the compliance of the film. In fact, if the size of the defect is comparable to the thickness of the film (i.e $a{\sim}H$), the pull-off stress (equation 5) immediately drops to a meaningful value: $\sigma_c \sim (W_a E / H)^{1/2}$. This co-operative development of flaws having a well-defined lateral spacing leads to some remarkable patterns in the adhesive film, which are discussed in the following sections.

## 3. Properties of Interfacial Elastic Instabilities

More than fifty years ago, Gent and Lindley [73] reported a path breaking experiment in which they observed sudden appearance of cavities (or cracks) in a rubber subjected to a tensile load. The size of the cavity and their spacing appeared to increase with the thickness of the rubber specimen. Later, Shull et al [74] observed that the free surface of an incompressible elastic solid can undergo a wavy undulation when it is subjected to a hydrostatic tension. Similar observation has also been made more recently [75] with soft hydrogels. All these instabilities are consequences of a material to preserve its volume when subjected to a hydrostatic tension. If such an experiment is performed with a low modulus viscoelastic adhesive against a surface with low adhesion, cavitation like instability is frequently observed at the interface, which progressively evolves into load carrying fibrils [44,49,50]. We show here that the development of such interfacial instabilities can be studied systematically with an elastomer where the viscoelastic loss is minimal.

As outlined above, the stressed elastic film gains compliance by forming multiple cavitation like bubbles at the interface. A slight perturbation on the surface thus allows the point of application of the stress to move, which reduces the potential energy at the cost of the increase of the elastic and surface energies. A cross-section through the interface would somewhat be like that shown in figure 3A, where part of the surface is in intimate contact and part is separated. This is a complex problem to solve mathematically as the solution of the elastic field equations ought to reproduce the correct shape of the surface corrugation as well. The problem is immensely tractable by considering a presumed shape of the surface in the form of a single sinusoidal mode, as shown in figure 3B, where the normal elastic traction is balanced by a van der Waals disjoining pressure and the Laplace pressure across the interface in its early stage. This picture of instability due to the competition of elastic and surface forces was already considered by Attard and Parker [76] in the context of interpreting the results of certain surface force measurements with deformable materials. This very basic idea was later translated in the subsequent works [30,31] in the field.

The picture of a principal periodic mode [77], as facile as it may be, nevertheless, gains support from an experiment in which the surface of a low modulus gel is examined under an optical profilometer soon after the contactor is removed (figure 3D). What one finds is that both the long and the short wavelengths of the surface features decline fast, but a particular sinusoidal mode decays most slowly. This is due to the fact that the energy of the system is concave with respect to the wavenumber ($k$), that has a minimum value corresponding to the principal mode (see below). Using a scenario that the overall energetics of the elastic deformation is controlled by this dominant sinusoidal mode, the dependence of the wavelength of instability on the thickness of the film can be demonstrated with a simple scaling analysis. In order to achieve this objective, we consider the elastic and surface energies (per unit area basis) in terms of the horizontal ($u$) and the vertical ($w$) displacements of the soft film as follows [33,41]:

$$\overline{U}_E + \overline{U}_S \sim \mu H \left( \frac{\partial u}{\partial z} + \frac{\partial w}{\partial x} \right)^2 + \gamma \left( \frac{\partial w}{\partial x} \right)^2 \tag{6}$$

This equation can be reduced further at the scaling level by choosing the characteristic length scales in the horizontal and vertical directions as the spatial wavelength ($\lambda$) and the thickness of the film ($H$) respectively. If the amplitude of the perturbation is $w_o$, the maximum horizontal displacement along the $x$ direction can be obtained from the equation of continuity $\left(\partial u/\partial x + \partial w/\partial z = 0\right)$, that is: $u/\lambda \sim w_o/H$ or $u \sim w_o\lambda/H$. Equation 6 can now be written as $\overline{U}_E + \overline{U}_S \sim \mu H w_o^2\left(\lambda/H^2 + 1/\lambda\right)^2 + \gamma\left(w_o^2/\lambda^2\right)$, which, upon minimization with respect to $\lambda$, yields the desired relation [41]:

$$\lambda \sim H\left(1 + \gamma/\mu H\right)^{1/4} \qquad (7)$$

When the ratio of the surface tension to elastic modulus is much smaller than the thickness of the film, one finds that the wavelength of instability depends only on the thickness of the film, i.e. $\lambda \sim H$. Vilmin et al [82] recently provided a scaling result of elastic instability ($\lambda \approx 4H$), by comparing the energy of forming a cavity of height $H$ with that required to create it by deforming the unstressed incompressible elastomer.

### 3.1. A Formal Analysis of the Dominant Wavelength of Instability

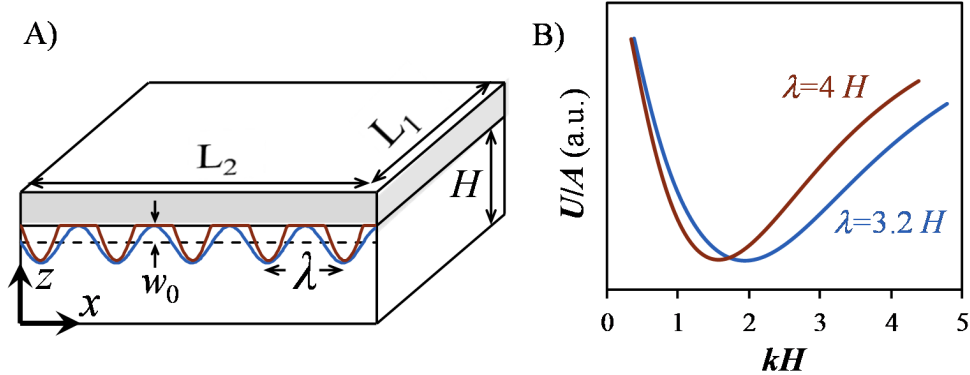

**Figure 4.** A) A schematic of a 1d model used to estimate the wavelength of instability. When a rigid contactor is pulled off a thin elastic layer bonded to a rigid substrate, its surface turns undulatory with portion of it remaining in contact with the contactor, while the rest debonding to form cavities. These cavities are considered cylindrical spanning uniformly along the length ($L_1$) of the film. B) The excess energy associated with the appearance of the instability varies non-monotonically with the dimensionless wave-number with a minimum at the dominant wave number. The blue curve represents the calculation based on a single mode of surface

perturbation. The maroon curve is estimated by considering that the deformation profile is a half rectified sine wave.

Figure 4A is a schematic of the surface of a thin soft film that has undergone an instability of wavelength $\lambda$ allowing a vertical displacement of the contactor as $w_o$. Ignoring the effect of surface tension and adhesion for the moment, we write down the sum of the potential and the elastic energies in the light of the general energy based approach of Kendall [79-82] as follows:

$$U = -\sigma_o w_o A + \frac{w_o^2 A M(k)}{4} \tag{8}$$

Where, the second term in equation 8 is obtained from the integral: $L_1 \int_0^{L_2} \int_0^{w_o} \sigma dl w dx$ (see figure 4A for the definitions of various terms) using an expression for the elastic stress that varies in proportion to the displacement as: $\sigma = \sigma_o + M(k)w_o \sin kx$. Here, $M(k)$ is the wavenumber ($k$) dependent stiffness of the film and $\sigma_o$ is the average stress, i.e. the applied external force ($F$) per unit area ($A$).

For a periodic perturbation of the surface ($w_o \sin kx$) profile, the deviatoric part of the stress is $M w_o \sin kx$. Substitutions of these expressions for the displacement and the stress in the Kerr's equation [70], the stiffness [71] of the film becomes:

$$M(k) = \frac{E}{H} \left[ \frac{1 - A_1(kH)^2 + A_2(kH)^4 - A_3(kH)^6 + A_4(kH)^8}{B_1 - B_2(kH)^2 + B_3(kH)^4 - B_4(kH)^6} \right] \tag{9}$$

All the coefficients ($A_i$, $B_i$) are functions of the Poisson ratio ($\nu$), which, are as follows:

$$A_1 = -\frac{1}{1-\nu}, \; A_2 = \frac{(3-4\nu)}{12(1-\nu)^2}, \; A_3 = -\frac{(3-4\nu)}{90(1-\nu)^2}, A_4 = \frac{(3-4\nu)}{1260(1-\nu)^2},$$

$$B_1 = \frac{(1+\nu)(1-2\nu)}{1-\nu}, \; B_2 = -\frac{(1+\nu)(3-4\nu)}{3(1-\nu)}, B_3 = \frac{(1+\nu)(3-4\nu)}{15(1-\nu)}, B_4 = -\frac{2(1+\nu)(3-4\nu)}{315(1-\nu)} \tag{10}$$

Using normal force balance condition, $\partial U / \partial w_0 = 0$, we can further simplify equation 8 as:

$$U = -\frac{\sigma_0^2 A}{M(k)} \tag{11}$$

Thus, only $M(k)$ has to be minimized in order to minimize $U$ (figure 4B), which yields the value of $kH$ as 1.95 with the minimum value of $M(k)$ as $2E/H$ for any applied load in the incompressible limit. Thus, the wavelength of the instability is: $\lambda \sim 3.2\,H$. While it correctly predicts that the wavelength of instability is proportional to thickness, the proportionality constant (3.2) is somewhat lower than that observed experimentally: $\lambda \sim 3.8\,H$ (figure 2). This discrepancy can, nevertheless, be reduced by considering a geometry of the instability closer to what is observed in experiments as discussed below.

## 3.2. An Improved Model

Although the analysis of the problem using the principal mode of surface undulation is a reasonable way to estimate the wavelength of instability, it does not quantitatively represent the underlying physics, as, in reality, part of the surface is in contact and part is not. We thus consider the surface corrugation as a half-rectified sine wave, and express the deflection of the surface and the corresponding stress in terms of the various Fourier modes as follows:

$$w = -w_0 \left[ \frac{1}{2}\sin(kx) - \frac{2}{\pi} \sum_{n=2,4,6}^{\infty} \frac{\cos(nkx)}{(n^2-1)} \right] \tag{12}$$

$$\sigma = \sigma_0 - w_0 \left[ \frac{M_1}{2}\sin(kx) - \frac{2}{\pi} \sum_{n=2,4,6}^{\infty} \frac{M_n \cos(nkx)}{(n^2-1)} \right] \tag{13}$$

We also consider the work of adhesion as comprised of two components, one is proportional to the area of contact and the other arising from the long range van der Waals forces as shown in the third and the second terms of equation 14:

$$\frac{U}{A} = -\frac{\sigma_0^2}{\left[\frac{M_1}{4} + \frac{4}{\pi^2} \sum_{n=2,4,6}^{\infty} \frac{M_n}{\left(n^2 - 1\right)^2}\right]} + \frac{(1-\phi)W_a}{2\pi} \int_0^{2\pi} \frac{d(kx)}{\left[1 + (w_0 / \ell_0)(1 - \cos kx)\right]^2} + \phi W_a \qquad (14)$$

Here, $\phi$ is the fraction of the total interfacial area that is in intimate adhesive contact. The last term of equation 14 certainly does not depend on $k$, whereas the second term becomes vanishingly small when the separation $w_0$ becomes much larger than the molecular distance of separation $\ell_o$. Each of the stiffness coefficients $M_i$ is a function of the wavenumber $k$, which can be evaluated using Kerr's expression (equation 9). Figure 4B (red curve) shows that the minimum value of U/A corresponds to $kH$=1.57, or, $\lambda = 4\,H$, which is now in much better agreement with the experimental observation: $\lambda = 3.8\,H$.

At this point, it should be mentioned that Kerr's model also allows for a finite friction to be considered at the rubber contactor interface. With a perfect no-slip condition at the interface, we find that $\lambda = 3.2\,H$. Thus, depending upon the value of the interfacial friction, the ratio $\lambda / H$ may vary from 3 to 4.

## 4. Cooperative Instability

In the absence of surface tension, the preceding analysis shows that the wavelength of instability is determined by the minimization of the longitudinal and transverse shear deformation energies of a single film. As the value of the elastic modulus itself is irrelevant in the minimization procedure, the wavelength depends only on the thickness of the film. The situation is, however, somewhat different for the case of the debonding of two soft films having dissimilar elastic moduli and film thicknesses [38]. We carry out this analysis at the scaling level by writing the total elastic energy (per unit area) of the two films as:

$$\overline{U}_{E,T} \approx \sum_{i=1,2} E_i H_i \left[\left(\frac{\partial u_{i,z}}{\partial x}\right) + \left(\frac{\partial u_{i,x}}{\partial z}\right)\right]^2 \ \text{ or, } \ \overline{U}_{E,T} \sim E_1 H_1 w_{01}^2 \left(\frac{1}{\lambda} + \frac{\lambda}{H_1^2}\right)^2 + E_2 H_2 w_{02}^2 \left(\frac{1}{\lambda} + \frac{\lambda}{H_2^2}\right)^2$$

$$(15)$$

Here, $E_1$ and $E_2$ are the elastic moduli of the films with thickness $H_1$ and $H_2$ respectively. This equation has to be minimized as before with respect to $\lambda$. However, as a common wavelength minimizes the energy of the composite system, the instability forms co-operatively that requires a condition for the continuity of the normal stress across the interface, i.e. $E_1 w_{01}/H_1^3 \sim E_2 w_{02}/H_2^3$ for $\lambda > H_i$. This condition along with $w_0 = w_{01} + w_{02}$ can be used to minimize $\overline{U}_{E,T}$ with respect to $\lambda$, which readily leads to an expression for the dominant wavelength of instability as:

$$\lambda \sim \left( \frac{E_2 H_1^7 + E_1 H_2^7}{E_2 H_1^3 + E_1 H_2^3} \right)^{1/4} \tag{16}$$

The right hand side of equation 16 can be treated as a normalized thickness ($\overline{H}$) of the composite film, which reduces to the result of a single film when the thicknesses of both the films are the same.

## 5. Pull-off Force

One of the main objectives for studying the morphological instability in thin film is to find out how it affects the pull off behavior of a rigid indenter from a soft confined film. In order to answer this question, we rewrite the total energy of the system as follows:

$$U \approx -\frac{F^2 H}{E(A_0 - A)} + W_a A \tag{17}$$

Where, $F$ is the applied force, $A_0$ is the total contact area and $A$ is the debonded (or the fluctuation) of the area of contact.

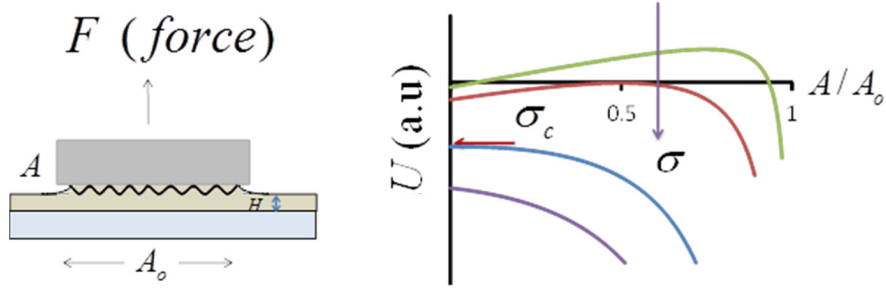

**Figure 5.** (Left) Schematic of an experiment in which a rigid block is being pulled off a thin elastic film bonded to a rigid substrate resulting in interfacial undulation. (Right) Total energy of the system as a function of the fractional area of contact depends on the applied stress [80]. The spinodal limit is attained at a critical stress when the rigid block is debonded.

Figure 5 shows that $U$ is a convex function of $A$, with an energy barrier preventing the spontaneous separation of the rigid block from the thin film. However, as the external force is increased, the energy barrier vanishes at a critical value of the force. This spinodal limit occurs when $A$ tends to zero, i.e. only a small fluctuation of the contact area is needed for fracture when the critical stress is reached. The condition of instability, thus, can be obtained from $(\partial U/\partial A)_{F, Lt\, A\to 0}=0$, [which also satisfies the unstable crack growth condition: $\partial^2 U/\partial A^2\,[=-W_a/(A_o-A)]<0$]. The stress needed to pull-off a rigid stud from a thin confined film, thus, is:

$$\sigma_c \sim \left(\frac{W_a E}{H}\right)^{0.5} \tag{18}$$

A similar analysis carried out for two dissimilar films leads to the following expression for the critical pull-off stress:

$$\sigma_c \sim \sqrt{\frac{W_a E_1 E_2 (E_1 H_2^7 + E_2 H_1^7)^{1/2}}{(E_1 H_2^3 + E_2 H_1^3)^{3/2}}} \tag{19}$$

For convenience, we write the above equation as:

$$\sigma_c \sim \left(\frac{W_a \overline{E}}{\overline{H}}\right)^{0.5} \tag{20}$$

## 5.1. Experimental Verification

The general equations requiring experimental verification are the linear dependence of the wavelength on the scaled thickness i.e. $\lambda \approx \overline{H}$ and that the critical pull-off stress to fracture increases with the square root of the ratio of the scaled elastic modulus to thickness, i.e. $\sigma_c \sim (\overline{E}/\overline{H})^{1/2}$. The fact that the pull off stress for thin confined film varies as $1/H^{1/2}$ was first demonstrated by Kendall [79] and later [44,45,74] by Creton, Shull, Crosby and their collaborators. The centro-symmetric pull off experiments of cylindrical block from the thin PDMS films as performed by Chung and Chaudhury [33] were mostly inspired by the previous works of Shull and Creton [45] -- both in terms of designing the specific apparatus and the method of measurement. However, in contrast to the prior studies, which mainly investigated the effect of film thickness on debonding stress, Chung and Chaudhury [33] extended their studies to films of different elastic moduli as well.

In order to carry out these experiments meaningfully, and thus to test the validity of equation 20 in terms of the role of elastic modulus, in particular, it is important to control numbers of variables. For example, a test of the equation $\sigma_c \sim (\overline{E}/\overline{H})^{1/2}$ assumes that the work of adhesion is independent of the molecular weight of the polymer, and thus its modulus. If the interaction between the two surfaces is not restricted to the mean field dispersion interaction, the effective work of adhesion would be amplified due to the stretching and relaxation [83-87] of the polymer chains (Lake-Thomas effect), in that $W_a \sim M_w^{1/2}$ or $W_a \sim E^{-1/2}$, $M_w$ being the inter-crosslink molecular weight of the polymer [88]. Furthermore, the interface could also age with time [86] when specific interactions are concerned, which can also be avoided largely by considering a pair of surfaces interacting via dispersion forces alone. It was therefore necessary to treat the rigid contactor with a low energy (methyl functional) self-assembled monolayer or by grafting polydimethyl siloxane chains upon it with high spatial density.

The soft elastomer was synthesized [89] from pure polydimethyl siloxanes without any added filler. Avoidance of commercial polymers is preferred as they contain various types of additives; furthermore, they are made of polymers having poly-dispersed molecular weights. We used carefully synthesized vinyl terminated dimethyl siloxane oligomers having relatively narrow dispersion of molecular weight [89]. These oligomers were crosslinked using a hydrido functional silane via a platinum catalyst to obtain elastomers of modulus ranging from 1 to 10 MPa, which exhibit negligible adhesion hysteresis with a –CH$_3$ terminated self-assembled monolayer. With these systems, two experiments were performed. In one experiment [33,38], a flat ended glass disc was brought into contact with a PDMS elastomer with a given modulus and thickness. After the contact was established, a computerized motion controller was used to separate the disc from the thin film while the debonding process was observed with a microscope (figure 6A). When a sufficient negative load was applied, instabilities developed in worm like patterns all throughout the contact (SI, Movie 1). At a critical load, the disc separated from the thin film abruptly. From the Fast Fourier Transform of the worm like patterns, the characteristic wavelength of the instability was estimated. These experiments were performed with a single and two films as shown in figure 6C. In each set of experiment, the pull-off force was recorded as a function of the modulus and the thickness of the PDMS film.

In a separate parallel experiment [32], a silane treated thin glass coverslip (or a cantilever) was brought into contact with the PDMS film by means of a spacer inserted at one end. Finger like instability patterns developed at the crack front, from which the wavelength of instability could be directly measured.

Figure 6B shows that the wavelengths of instability for both the single and two film cases follow the relationship: $\lambda = 3.9\overline{H}$. Similarly, the critical stress needed to separate either a bare or a thin film coated rigid flat substrate from another thin film coated rigid support also varies linearly with the scaled ratio of the elastic modulus and the thickness of the composite films. Using a work of adhesion as 40 mJ/m$^2$, we obtain the experimental relationship between the pull-off stress and the scaled ratio of the elastic modulus and the film thickness as: $\sigma_c = 1.9(W_a\overline{E}/\overline{H})^{1/2}$. The prefactor (~ 2) is about twice of what is expected from this two dimensional analysis. We expect that that a proper 3d analysis might be able to reduce the gap of disagreement. However, since the correction factor is on the order of unity, we might conclude that the agreement between theoretical analysis and experimental results is rather good.

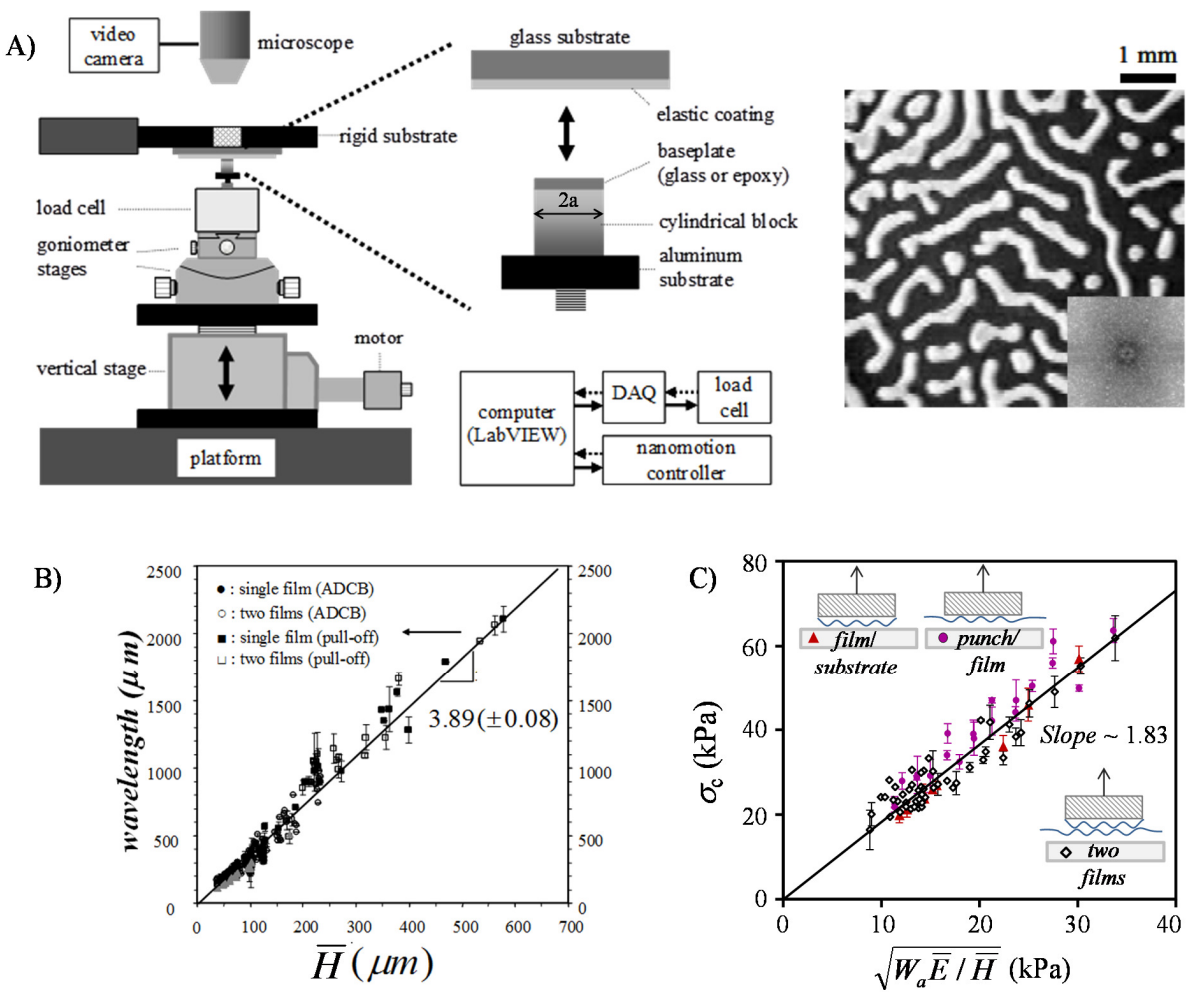

**Figure 6.** (A) Schematic of the apparatus [33] used to measure pull-off forces. These experiments were carried out in three ways: (i) a rigid punch is pulled off an elastic layer bonded to a another rigid substrate, (ii) a rigid punch coated with an elastic layer is pulled off a rigid uncoated substrate, (iii) the rigid punch coated with an elastic layer is pulled off an elastic layer bonded to another rigid substrate. (B) The wavelengths of instabilities in all cases cluster around a single line: $\lambda = 3.89\overline{H}$. (C) The critical stress of separation scales linearly with $\left(\overline{E}/\overline{H}\right)^{1/2}$. (A is reprinted with permission from Ref. [33]. Copyright (2005) Taylor & Francis Inc. B is reprinted with permission from Ref. [38]. Copyright (2006) The European Physical Journal).

We note that the scaling results obtained for the pull-off of a rigid block from a thin film also applies when the former is de-bonded from the soft film in a shear mode [40,90]. In these studies, the surface of the rigid block was modified either with a hydrophobic silane comprising of methyl groups (to ensure dispersion interaction) as in the pull off tests or with alternate layers of Poly(allylamine hydrochloride) (PAH) and poly(acrylic acid) (PAA) to introduce specific interactions. The multilayer film was formed on a PDMS surface in such a way that its first layer is PAH, whereas its last layer interacted strongly with either a PAA or PAH terminated multilayer of the counter surface. With such a design, when the rigid block is placed against the PDMS film and removed after an hour, the polyelectrolyte multilayer assembly is always delaminated from the PDMS surface [90]. Thus, the measured strength is always that of the polyelectrolyte layer, or a –CH3 terminated monolayer, with the PDMS film.

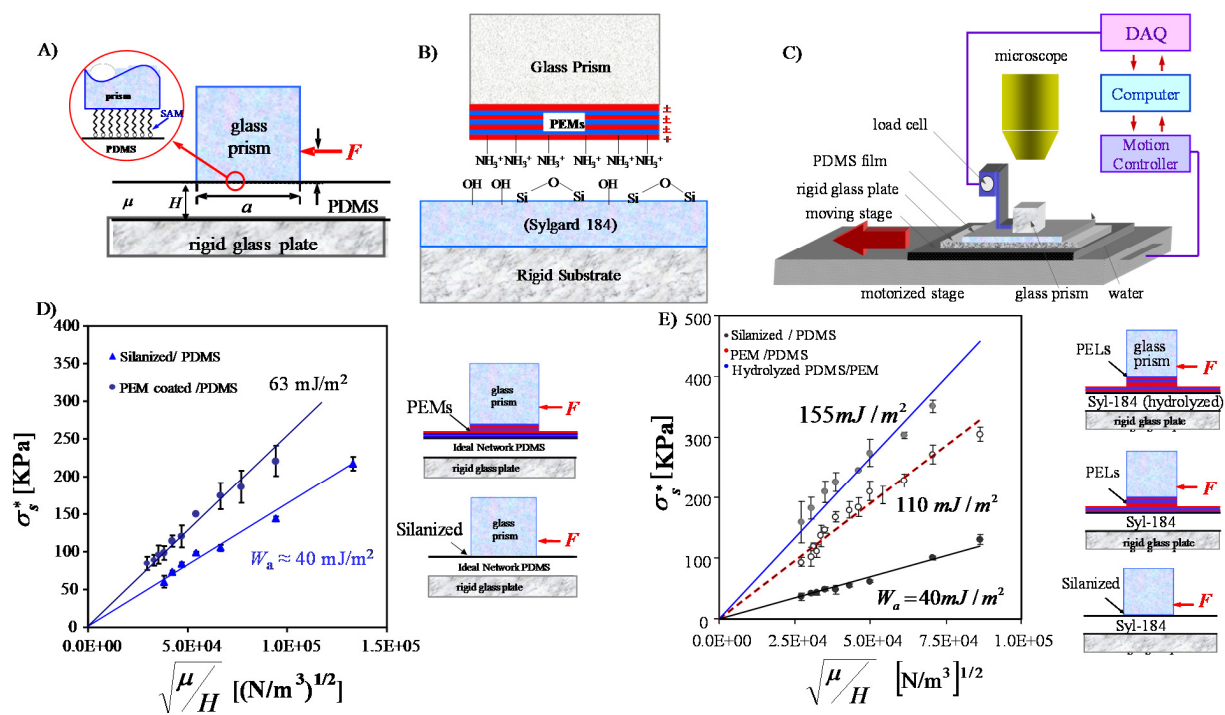

**Figure 7.** (A) Schematic of the contact of thin PDMS elastomer and a silane coated glass disc. (B) A polyelectrolyte multilayer with terminal NH3+ groups could interact weakly with the siloxane or any residual hydroxyl groups of a PDMS surface. The polyethylene multilayer, itself, is formed by sequential depositions of positively charged Poly(allylamine hydrochloride) (PAH) and negatively charged poly(acrylic acid) (PAA). (C) Schematic of the apparatus [90] used to measure the critical force of debonding. (D-E) Plots show that the critical shear stress of debonding for various experiments vary inversely with the square root of the film thickness. (Reprinted with permission from Ref. [90]. Copyright (2009) Taylor & Francis Inc.)

The data summarized in figures 7 C-D show that the critical shear stress of debonding for an ideal PDMS network and a commercial polymer (sylgard 184) against a silanized glass cube also varies inversely with the square root of the thickness of the film. Both the elastomers, however, interact more strongly with a polyelectrolyte containing $NH_3+$ groups as compared to a surface modified with a $-CH_3$ terminated self-assembled monolayer. This difference is attributed to the fact that while both Sylgard 184 and an ideal PDMS network interact with a $-CH_3$ functional surface via the dispersion force alone, they also participate in some specific (i.e. donor-acceptor) interactions with an -$NH_3+$ functional surface. This interaction is, indeed, much stronger for a hydroxyl functional hydrolyzed PDMS in contact with an $NH_3+$ functional surface.

## 6. Elasto-Capillary Phenomena: when Surface Tension Dominates

So far, we discussed how the wavelength of instability of a single confined film depends only on its thickness whereas for two elastic films, the wavelength does depend on their material properties due to the co-operative energy minimization. We show here that even for a single film, $\lambda$ can depend on the material properties if it is so soft and/or so thin that the role of surface tension cannot be neglected. Such a supposition has already been evident in the scaling relation: $\lambda \sim H \left(1 + \gamma/\mu H\right)^{1/4}$, according to which, $\lambda$ is proportional to $H$ when $\gamma / \mu << H$. There can be a regime when $\gamma / \mu$ is only slightly less than $H$, so that the wavelength is an additive function of the film thickness and the elasto-capillary length, i.e. $\lambda \sim H + \gamma / 4\mu$. On the other hand, if $\gamma / \mu >> H$, the elasto-capillary and the geometric length scales cannot be separated in a trivial way as the wavelength of instability, $\lambda \sim (\gamma H^3/\mu)^{1/4}$.

These long wavelength dominated pattern formation can be easily visualized with a thin

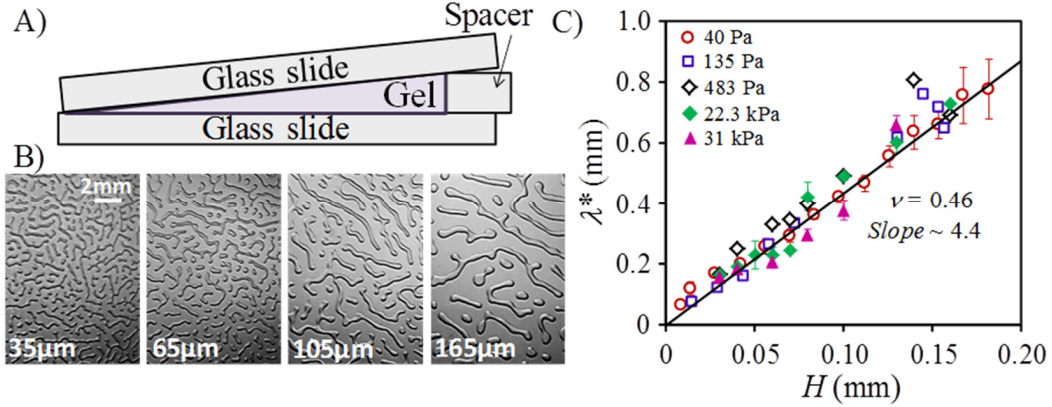

**Figure 8.** Effect of surface tension on instability. (A) A hydrogel film inside a wedge formed by two inclined glass plates [77]. After the solution is gelled, the upper glass plate is slightly peeled off from the thicker side of the hydrogel film in order to induce the instability. This gradient gel allows measurement of the wavelength of instability as a function of a wide range of film thicknesses in a single specimen. (B) Worm like patterns at the interface of the gel and the top plate. (C) The scaled wavelengths $\lambda^* = \lambda/(1+\gamma/\mu H)^{1/4}$, as estimated from equation 21, vary linearly with the film thickness ($H$). The experimental wavelengths are rescaled as $\lambda^* = \lambda/(1+2\gamma/\mu H)^{1/4}$. (A & B are reprinted with permission from Ref. [77]. Copyright (2013) The American Chemical Society)

hydrogel film with $\gamma/\mu \sim 1$mm. Indeed, the wavelength of instability observed with such a film of $\mu \sim 100$ Pa is seven to eight times that of its thickness, which contrasts the $\lambda = 3.8H$ behavior observed with higher modulus films (figure 2).

We investigate here a formal relationship between $\lambda$ and $H$ using a linear stability analysis of energy of the deformed film similar to that used earlier, but, at this juncture, adding the additional stiffness of the film contributed by its surface tension:

$$\frac{U}{A} = -\frac{\sigma_0^2}{\varkappa k^2 \left[ \frac{1}{4} + \frac{4}{\pi^2} \sum_{n=2,4,6}^{\infty} \frac{n^2}{(n^2-1)^2} \right] + \left[ \frac{M_1}{4} + \frac{4}{\pi^2} \sum_{n=2,4,6}^{\infty} \frac{M_n}{(n^2-1)^2} \right]} + \phi W_a \tag{21}$$

The energy $\overline{U}$ (= U/A) of the system is minimized by varying $kH$ for $\mu$ (shear modulus) ranging from 40 Pa to 33 kPa, but by fixing the surface tension of the hydrogel to a value of 72 mN/m.

The numerical method is same as above (see equation 9-13), except that the calculations were performed with a Poisson ratio of 0.46, in view of what has been reported in the literature for a polyacrylamide hydrogel film [91]. These calculations reveal that there is an interpolating relationship $\lambda \approx 4.4H(1+\gamma/\mu H)^{1/4}$ that connects the behavior of the film from the small to a large capillary number limit, which is consistent with the scaling relation: $\lambda \sim H (1+ \gamma/\mu H)^{1/4}$, as discussed above. We now compare and contrast the above interpolation equation with those observed in the following experiments.

When the experimentally observed wavelength of instability is scaled by dividing it with $(1+2\gamma/\mu H)^{1/4}$ and plotted against the thickness $H$, all the data, indeed, cluster around a straight line (figure 8C). Note that there is a difference of the prefactor $\gamma/\mu H$ in experiment and theory. Both the equations, nevertheless, lead only to a small discrepancy between the experimental [ $\sim 5.2(\gamma H^3/\mu)^{1/4}$] and the theoretical [$\sim 4.4(\gamma H^3/\mu)^{1/4}$] value for $\lambda$ in the capillary dominated limit. These expressions for the long wave limit of the instability can be compared with a simple analysis for an incompressible ($\nu$=0.5) material using the principal mode $k$, which can be obtained by minimizing the stiffness of the film as $M^*(k) \sim E/(k^2 H^3)+\gamma k^2$ , where the first term of the right hand side is obtained from the limiting expression of $M(k)$ (equation 9) for $kH$ <<1. Minimization of this stiffness yield an expression for the wavelength as $\lambda \sim 4.8(\gamma H^3/\mu)^{1/4}$ , which is, indeed, very similar to the relations found in experiment and detailed analysis as reported above. We point out here that Gonuguntla et al [36] too obtained essentially the same equation previously for an elasto-capillary instability.

### 7. Role of Elasto-capillarity in Adhesive Fracture: Modified Griffith Equation

The long wave instability observed with low modulus films also has an important implication on the stress ($\sigma$) needed to pull off a flat ended rigid stud from such a film as the governing equation 17 (using the dominant wavelength of instability) is to be modified as follows [77]:

$$U \sim -\frac{F^2}{\left(\gamma k^2 + M\right)\left(A_0 - A\right)} + \phi W_a A \tag{22}$$

Since, $M$ is on the order of $\gamma k^2$ (see above), we can also rewrite equation 22:

$$U \sim -\frac{F^2}{\left(2\gamma k^2\right)\left(A_0 - A\right)} + \phi W_a A \tag{23}$$

Now, using the longer wave limit of the instability, i.e. $k^{-1} \sim (\gamma H^3/E)^{1/4}$ and using the usual fracture criteria: $\left(\partial U/\partial A\right)_{F, Lt\, A\to 0} = 0$ [which also satisfies $\partial^2 U / \partial A^2 < 0$], we obtain a new expression for the pull-off stress as:

$$\sigma_c \approx \left(\frac{\gamma W_a^2 E}{H^3}\right)^{\frac{1}{4}} \quad \text{or} \quad \sigma_c \approx \left(\frac{\gamma}{EH}\right)^{0.25}\left(\frac{W_a E}{H}\right)^{0.5} \tag{24}$$

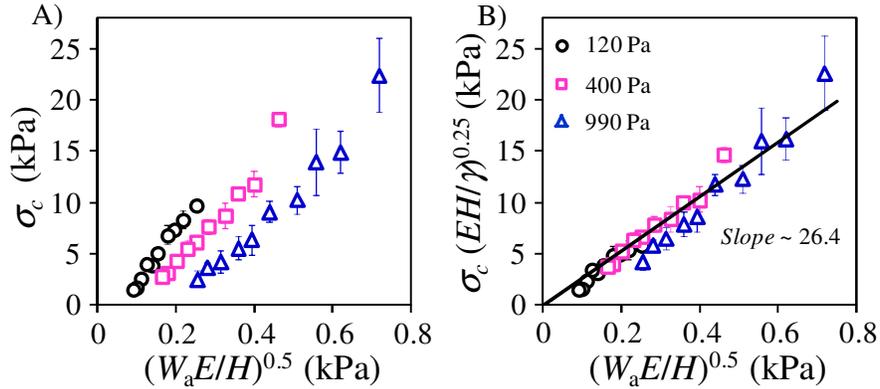

**Figure 9.** A) The normal stress to debond a silanized circular rigid glass disc (12.8 mm diameter) from hydrogel films of different thickness and modulus are plotted against $\left(W_a E/H\right)^{0.5}$. B) The data of figure A cluster around a single line when the stress is multiplied by $\left(EH/\gamma\right)^{0.25}$. (These experiments were performed with Chih Hsiu Lin).

Thus for the ultra-soft and/or very thin films, the normal stress required to detach a flat ended rigid block from a thin confined film is expected to deviate from the Griffith relationship (equation 18) due to the additional multiplicative term: $(\gamma/EH)^{0.25}$.

While the measured pull-off stress of a rigid stud from the hydrogel films of different shear moduli and thicknesses show that the data conform to the scaling relationship shown in equation 24, the slope of the plot $\sigma_c(EH/\gamma)^{0.25}$ versus $(W_a E/H)^{0.5}$ is substantially greater than unity. It is important to note that although a certain level of amplification of the Griffith's stress is expected if $\gamma/EH$ is substantially greater than unity, its magnitude is not high enough in the current study that would explain the discrepancy. While we do not have a definite explanation for this result, we speculate on certain factors that might be considered in future for an amicable explanation of the discrepancy in the summary section.

## 8.  Fingering (Saffman Taylor like) Instability in Thin Elastic Film

What is intriguing with the adhesion induced elastic instability is that it displays various analogs of other well-known instabilities [92-95] in viscous liquids and viscoelastic polymers, including the saw-tooth instability [96] observed in polymer processing. Therefore, an in-depth, or even a cursory, understanding of the properties of the elastic instabilities may be useful in gaining insights into how similar features evolve in more complex situations.

We remind the readers that we already discussed an elastic analog of the canonical Saffman-Taylor fingering instability in section 5.1, which develops at the crack front when a thin flexible plate is separated from a thin elastomeric film in a cantilever geometry. Experiments [97-100] show that the spacing of the (elastic) fingering instability, in general, follows: $\lambda \approx 4\overline{H}$, which reduces to the result $\lambda \approx 4H$ for a single film. For the latter case, the wavelength of instability is not only independent of the elastic modulus of the film, but it is also independent of the bending rigidity of the cantilever and even the work of adhesion. The amplitude of the instability, in contrast to the wavelength, however, increases rather nonlinearly (Figure 10) with various material parameters. These fingering patterns appear almost instantaneously when a film of thickness less than a critical value is brought into contact with the plate and they remain so irrespective of whether the contact line propagates or stops moving [98]. In this respect, these U-shaped patterns (figure 10) are vastly different from that of the Saffman-Taylor instability, even though they have deceptively similar appearances. These fingering patterns develop only when the thickness of the film is below a critical value [29,31] with an amplitude ($A_m$) that increases with the rigidity of the cantilever , i.e. the degree of confinement (see also the next section), but it decreases (figure 11B) with the shear modulus of the film: $A_m \propto (D/\mu)^{1/3}$. The control parameter ($\varepsilon$) governing the amplitude of instability can be written in a non-dimensional form as the ratio of the film thickness ($H$) and the stress decay length [$\beta = \left(DH^3/3\mu\right)^{1/6}$] along $x$ direction (figure 11), i.e. $\varepsilon = H/\left(DH^3/3\mu\right)^{1/6}$. The instability ensues when these two length-scales remain separated by at least a factor of 0.3, i.e. $\varepsilon_c = H_c/\left(DH_c^3/3\mu\right)^{1/6} = 0.3$. In most situations, however, $\varepsilon$ remains significantly smaller than 0.3. This is a 3d problem where a linear analysis is insufficient to fully characterize the properties of the ensuing instability. Here, the evolution and the fine structures of the morphology are driven by the minimization of the energy comprised of the bending of the contacting plate, the shear deformation of the film, the adhesion between the plate and the film, and the surface energy associated with the creation of curved surface near the finger region. This last contribution, however, is usually negligible in comparison to the elastic energy for a polymer of modulus in the range of 1MPa., i.e. when the elasto-capillary number $\gamma/\mu H$ is much less than unity. An expression of the total energy of the system is:

$$\Pi = \Pi_{elastic(film)} + \Pi_{bending} + \Pi_{adhesion} \quad = \frac{\mu}{4}\int_{-\infty}^{0}\int_{0}^{2\pi/k}\int_{0}^{H}\left(\left(v_z+w_y\right)^2+\left(u_y+v_x\right)^2+\left(u_z+w_x\right)^2\right)dzdydx +$$
$$\frac{D}{2}\int_{-\infty}^{b}\int_{0}^{2\pi/k}\left(\psi_{xx}\right)^2dydx + W_A\left(\frac{2\pi b}{k}+A_{finger}\right)$$

(25)

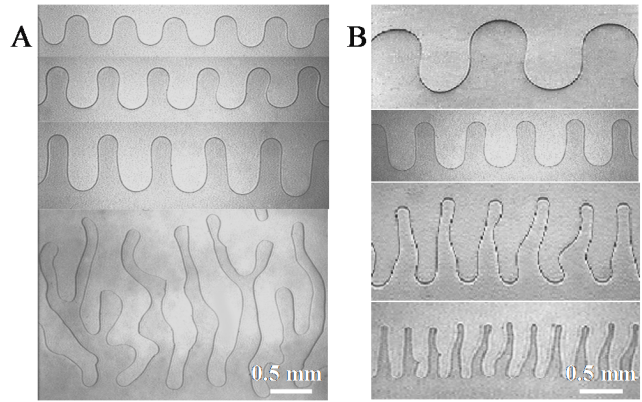

**Figure 10.** Optical micrographs of fingering patterns [29] that develop in the contact line when a thin flexible plate is removed from a thin PDMS film. Glass cover slips of flexural rigidity $D$ ranging from, 0.02 to 1.35 Nm are lifted off films of thickness $H$ ranging from 40 to 350 µm and shear moduli ($\mu$) ranging from 0.25 to 2.0 MPa. Micrographs in A) correspond to $H = 150\,\mu m$, $\mu = 1.0\,\text{MPa}$ and $D = 0.02, 0.09, 0.2$ and 1.0 Nm respectively (top to bottom). Micrographs in B) correspond to ($H = 40\,\mu m$, $\mu = 2.0\,\text{MPa}$ and $D = 0.02\,\text{Nm}$), ($H = 80\,\mu m$, $\mu = 0.25\,\text{MPa}$ and $D = 0.02\,\text{Nm}$), ($H = 160\,\mu m$, $\mu = 1.0\,\text{MPa}$ and $D = 0.2\,\text{Nm}$) and ($H = 350\,\mu m$, $\mu = 0.25$ MPa and $D = 0.4\,\text{Nm}$) respectively (top to bottom). (A is reprinted with permission from Ref. [29]. Copyright (2000) The American Physical Society. B is reprinted with permission from Ref. [32]. Copyright (2003) The American Chemical Society.)

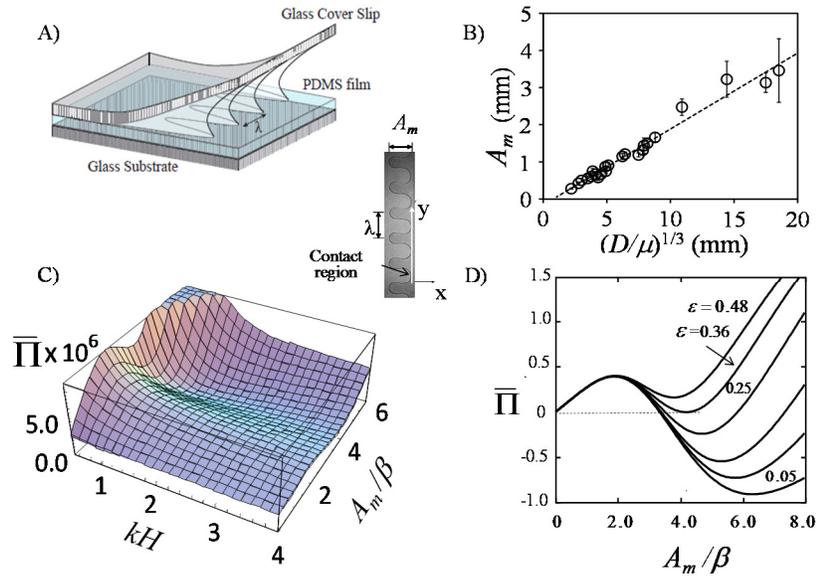

**Figure 11.** (A) Schematic of the contact line region of the thin film, when a flexible plate is peeled off a thin PDMS film bonded to a rigid substrate (Redrawn from [78]). (B) Experimentally measured amplitudes of instability as obtained from various experiments are linear with $(D/\mu)^{1/3}$. (C) bifurcation diagram [98] of the total excess energy of the system as a function of the amplitude and the wavenumber of the instability. (D) Variation of the excess energy with the amplitude of instability at different values of $\varepsilon$. (C & D are reprinted with permission from Ref. [98]. Copyright (2006) The American Physical Society).

Where, $u$, $v$ and $w$ are the deformations in the film along $x$, $y$ and $z$ directions respectively, $\psi$ is the vertical displacement of the plate from undeformed surface of film, $A_{\text{finger}}$ is defined as the interfacial area of contact at $-b < x < 0$; and $k$, as usual, defines the wave number of instability. The above expression is simplified by assuming that the contacting plate bend only in the direction of propagation of the contact line (i.e. along $x$) while remaining uniform along the wave-vector (i.e. along $y$). During the peeling action on the cantilever, the elastomeric film deforms normal to the surface, with a concomitant depression close to, but behind, the contact line. For a thick film, the shear deformation occurs in both the $xy$ and $yz$ planes i.e. in planes normal to the $z$ and the $x$ axes respectively. Such a deformation causes depression in the film ahead of the line of contact of the film and the plate, which compensates the normal deformation of the film. When this compensation is of the right amount, no undulation needs to be formed in the contact line region. However, for sufficiently thin films, the shear deformations also occur in the $xz$ (i.e. normal to $y$ axis) plane. It is this additional lateral deformation that results in the undulation of the contact line, which is accounted for in the energy calculations. The remaining component of the energy is the work of adhesion, which depends on the area of contact between the film and the cover plate. There is, thankfully, a geometric non-linearity in the current problem as the overall energy is to be minimized with respect to both the wavelength and the amplitude. This non-linearity plays a vital role in generating the bifurcation diagram (figure 11 C) of the morphology of fingers in which the total energy goes through a minimum for a given value of the parameter $\varepsilon$.

While equation 25 defines the total energy of the system, appearance of the instability is best described by considering the excess energy over and above that of the unperturbed homogeneous deformation of the film: $\Pi_{excess} = \Pi - \Pi_0$. This excess energy is a function of three different sets of parameters: the control parameter ($\varepsilon$), the amplitude $A_m$, wave number $k$; and the geometric lengths of the experiment, i.e. the crack length $b$ and the spacer height $\Delta$. The instability ensues with finite amplitude when these parameters are such that the dimensionless excess energy $\overline{\Pi}\left(\equiv \Pi_{excess} / \mu\beta^3\right)$ of the system, varying non-monotonically with the amplitude ($=A_m/\beta$), becomes negative (figure 11D) . It is easy to appreciate that the contact line remains straight so long as the minimum of the excess energy remains positive, however, it turns wavy as $\Pi_{\min}$ becomes negative. In this problem, $\overline{\overline{\Pi}}$ remains positive at any value of $A_m$ when $\varepsilon = 0.48$, implying that the undulation of a straight contact line would not occur as it would increase the total energy of the system. On the other hand, $\varepsilon = 0.36$ presents a limiting case for which minimum of the excess energy $\Pi_{\min}$ attains zero; while its value remains negative for $\varepsilon < 0.36$. Taken together, this analysis suggests that the contact line becomes unstable when the control parameter $\varepsilon$ is less than a critical value, $\varepsilon < \varepsilon_c = 0.36$. In reaching the final amplitude corresponding to the minimum excess energy starting from zero amplitude may not always be possible as there exists an energy barrier (see the next section further comments on this subject).

From figure 11C, it is evident that $\overline{\overline{\Pi}}$ attains a minimum for $kH = 1.91$. The wavelength of perturbations thus scales with the thickness of the film as $\lambda = 3.3H$, which is similar to the general observations in a wide range of experimental geometries. The theoretical analysis, thus, captures the main points of our experiments in that (a) the amplitude increases with the increase in confinement of soft film; (b) undulations of the contact line ensues when the film becomes more than critically confined, i.e. $\varepsilon < 0.36$; (c) only perturbations with finite amplitude grow, while the others decay. The energy barrier renders this elastic instability different from other similar instabilities observed in liquid systems. A super-critical Saffman-Taylor instability, nevertheless, has been observed in viscoelastic liquids; that again is due to its elastic property [101].

We now return to the important issue raised above that an energy barrier has to be overcome in order for the system to reach its globally minimal energy state and that for an under-critically confined film no fingering instability is possible. All these considerations also point out that the amplitude of undulation never increases gradually from zero, but it appears abruptly at a finite value, which, in general, is consistent (see also section 9) with the experimental observations [29,32]. The question, nevertheless, remains as how it may be possible for the system to reach the minimum energy state even for the super-critically confined films as an energy barrier always has to be overcome.

Part of the answer to this question lies in the fact that the fingering patterns evolve from another state, in which bubbles cavitate at the onset of the separation of the cantilever from the film. Figure 12 summarizes the result of a typical displacement controlled peel experiment [34,41], which is carried out in a style similar to that of Wan et al [102-104] , who studied adhesion and fracture with cleaved mica. In our experiment, one end of the cantilever is lifted from a thin PDMS film with the help of a computerized motion controller and the contact line region is observed with a microscope. One of the main observations in this study is that bubbles cavitate [34,41] slightly inside the edge of contact, which corresponds to the region of maximum tensile stress (figure 12A). The cavities eventually grow and coalesce with each other forming two cracks, which propagate in opposite directions. While one of the cracks propagates along the length of the undetached cantilever-PDMS interface, it already has the incipient fingering pattern, which becomes fully developed soon after the other crack reaches the edge of contact. At this point, the cantilever detaches from the edge of the PDMS film rather abruptly. There is, therefore, a sudden release of the system's energy, which is powerful enough to undulate the contact line further, thus, overcoming the energy barrier of the type reported in figure 11.

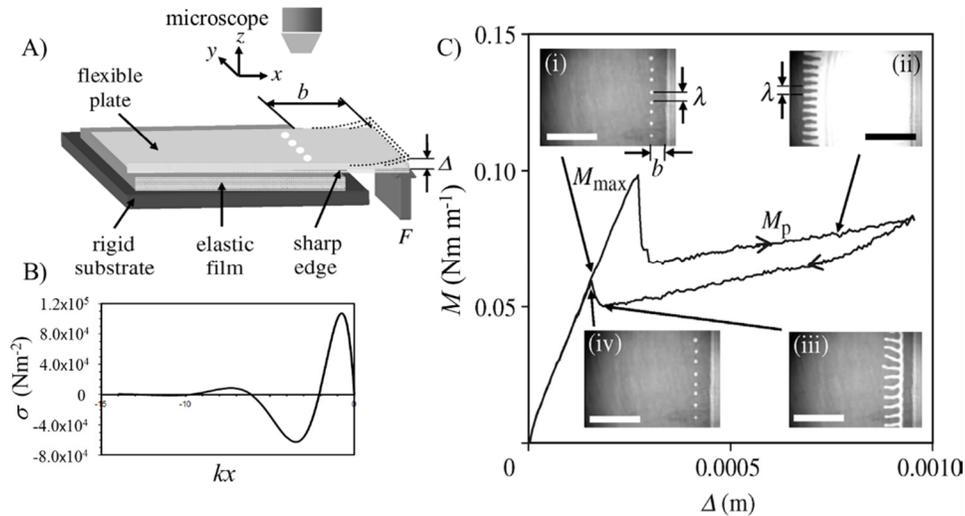

**Figure 12.** (A) Schematic representation of an experiment [41] in which a flexible plate is detached from the layer of a rubber film, by applying a lift-off load $F$ at its free end. (B) The normal stress is damped oscillatory as a function of distance from the edge of the film. The bubble like cavitation is observed where the tensile stress is maximum (C) The applied torque $M$ over a cycle of detachment and attachment of the flexible plate on a thin PDMS film shows hysteresis. Evolutions of the cavitated bubbles are shown in the inset. Scale bars in C) are 3 mm. (Reprinted with permission from Ref. [41]. Copyright (2005) The Royal Society)

## 9. Supercritical Instability

While morphological instability appears in various systems, it is not always easy to study systematically how its amplitude evolves in terms of a control parameter that triggers it in the first place. For example, the energy of the system for a flat ended rigid stud in contact with a confined thin film is convex with respect to the area of contact. There, the instability does not develop till the critical condition of fracture is approached. This situation, thus, leaves only a small window of opportunity to study the onset of instability as the fracture proceeds soon after the instability develops. The total energy of a cantilever in contact with a thin elastomeric film is, however, concave with respect to the crack length. It is thus possible to advance and recede the crack (as is the case with the contact mechanics measurements of spherical contacts) quasistatically without worrying about an imminent instability of contact [99].

The particular experiment that allows a study of the onset and the evolution of the instability is with a thin cantilever (a glass cover slip) placed on the PDMS film with two spacers placed on the opposite sides of contact. Here, the meniscus instability develops at both the crack fronts, which can be manipulated by changing the distance between the spacers as a function of the bending rigidity of the plate as well as the thickness and the modulus of the PDMS film.

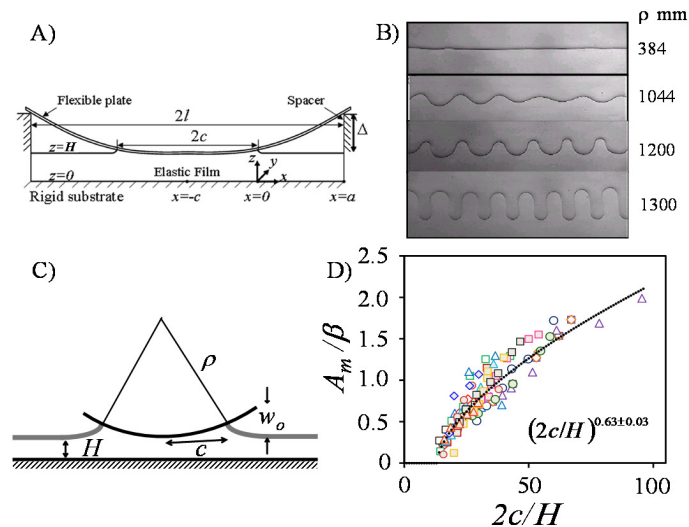

**Figure 13.** (A) A flexible adherent is lifted off a thin layer of adhesive in double cantilever [29,99] geometry by using two spacers inserted at its two sides generating two wedge cracks which tends to propagate towards each other. (B) Optical micrographs of the instability pattern along a contact line for different curvatures of the contactor. (C) Schematic of the side view of the plate in contact with the film (D) Dimensionless amplitudes of the fingers obtained from large number of experiments are plotted against the confinement parameter defined as the ratio of the width of the contact and the film thickness. (A is reprinted with permission from Ref. [99]. Copyright (2007) Taylor & Francis Inc.)

The analysis in the preceding section already indicated that a second order phase transition of the pattern formation is possible with respect to the control parameter $\varepsilon$. This fundamental problem can be studied more rigorously using the current method by varying $\varepsilon$ systematically. Experimentally, it is very clear that the amplitude of the instability depends on the width of the contact, or, operationally, the distance between the spacers. Under a small bending approximation, the width of the stressed zone $2c$ scales as $2c \sim (w_o \rho)^{1/2}$, where $w_o$ is the vertical displacement of the elastomer near the crack tip and $\rho$ is the radius of curvature of the plate. However, as fracture mechanics dictates that $w_o \sim (W_d H/\mu)^{1/2}$, we expect the amplitude to vary with the radius of curvature as: $A_m \sim \rho^{1/2}$. Noting that the width $2c$ is the characteristic length-scale, the control parameter for this experiment can be defined as $\varepsilon = H/2c \sim H/(w_o \rho)^{1/2}$. When the radius of curvature is small, no instability develops as the film is poorly confined. Conversely, for large radius of curvature of plate, $\varepsilon$ diminishes (i.e. high confinement), which is expected to render the two sides of the contact width unstable to finite perturbations. This simple picture illustrates that a supercritical instability with respect to the degree of confinement may be observed if the radius of curvature is varied systematically. Such a behavior is illustrated in figure 13 where the amplitude (non-dimensionalized by the stress decay length) is plotted against $2c/H$ -- the degree of confinement. These data obtained with 15 sets of experiments using different combinations of the modulus and the thickness of PDMS and the flexural rigidity of the cantilever show that the amplitude of the instability remains zero below a threshold limit of $2c/H$, beyond which it grows as the confinement increases.

It is to be noted that fundamentally both the experiments, one presented here and the one depicted in section 8 are similar, as both explore the effect of confinement of a thin elastic film subjected to contact stresses, nevertheless, with a difference. While the characteristic length scale $\beta$ along $x$ is chosen by the system via balance of forces in the experiments of section 8, it is imposed upon the system in the experiments described in this section.

## 10. Elastic Engulfment

The competing forces arising from the bending of the flexible plate, the deformation of the confined film and the work of adhesion between the two surfaces often give rise to some counter-intuitive situations. For example, let us consider an experiment that is an inverted image of that reported above in the sense that the central part of the plate is detached from the film by means of a thin wire (20 μ*m* diameter), but it is, otherwise in contact with the rest of the film.

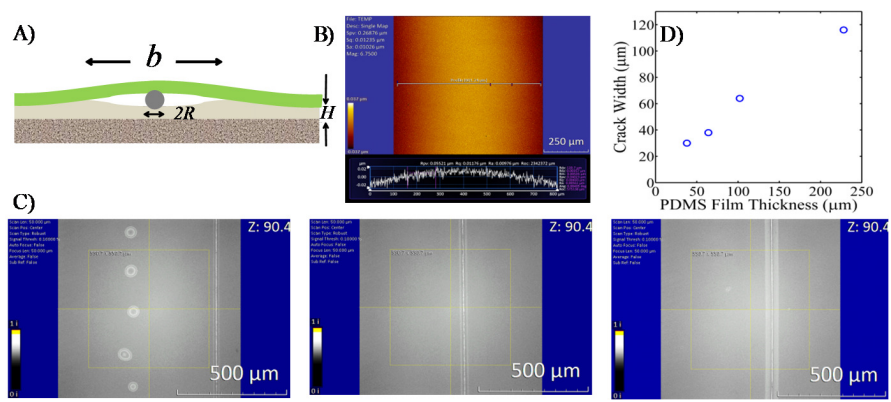

**Figure 14.** (A) Schematic of an experiment in which a flexible glass cover slip is brought in contact with a thin elastomeric film (*E*~3 MPa) over a 20 μm diameter tungsten wire. (B) A surface profilometric image of the plate above the wire shows the bending of the plate (C) (i) Interfacial bubbles form far from the wire for a 38 μm film decays slowly with time as seen in micrograph (ii) which is captured after 4days of making the first contact. Micrograph (iii) demonstrates the crack opening around the wire over a 102 μm film that was also captured after 4 days. (D) the crack widths surrounding the wire is roughly linear with the thickness of the film. (These experiments were performed by Jonathan E. Longley. The original observation of this effect was made along with A. Jagota).

With all these films, a crack like gap develops around the wire, the width of which increases with the thickness of the film. For the thinnest film (38 μ*m*) , the gap appears to close soon after the contact is made; but, bubble like instability becomes visible far from the wire at the interface of the plate and the film. These bubbles, however, disappear slowly with time and a more pronounced gap manifests in the central region of contact. Figure 14 shows that the width of the crack at equilibrium varies linearly with thickness. Even in the absence of a detailed analysis, the physical picture of the problem is revealed if the width of the crack is taken to be half of the wavelength of instability that minimizes its longitudinal and transverse shear strain energies of the film. The width of the crack should thus be proportional to the thickness of the film as is observed experimentally.

## 11. Analog of Rayleigh Instability in Elastic Interface

Morphological characteristic of the Rayleigh instability [1] is that a thin liquid thread breaks up into well-spaced small droplets, the spacing of which is independent of its material property, i.e. the surface tension, which triggers the instability in the first place. Here, we show that this very feature of Rayleigh instability can also be observed in confined elastic [34,41] film whenever a long air channel is trapped at the interface of a thin soft film and a contactor. One example of such an instability is evident in the healing of a blister (figure 15A), where air is trapped in long channels resulting from an elastic instability (Appendix B, Movie 2). These hydraulic channels are the escape routes for the trapped air with a corresponding Darcy's permeability ($k_D$) that strongly depends upon the hydraulic diameter of the channels and thus the thickness of the soft film as $k_D \sim H^{1.5}$. Here, the interfacial adhesion closes the channel, but the pressure of the air inside the channel and the elastic energy of the film increase. The system relieves the constraint by undergoing a morphological transition, in which the long channel breaks up into bubbles. Another example of such an instability is related to the experiment reported in section 8, where we discussed the sequences of events encountered during the separation of a cantilever from the film. After the crack advances to a certain degree, if its direction of motion is reversed by lowering the end of the cantilever, the force supporting the crack follows a slightly different path, indicating that there is a certain amount of adhesion hysteresis in this problem.

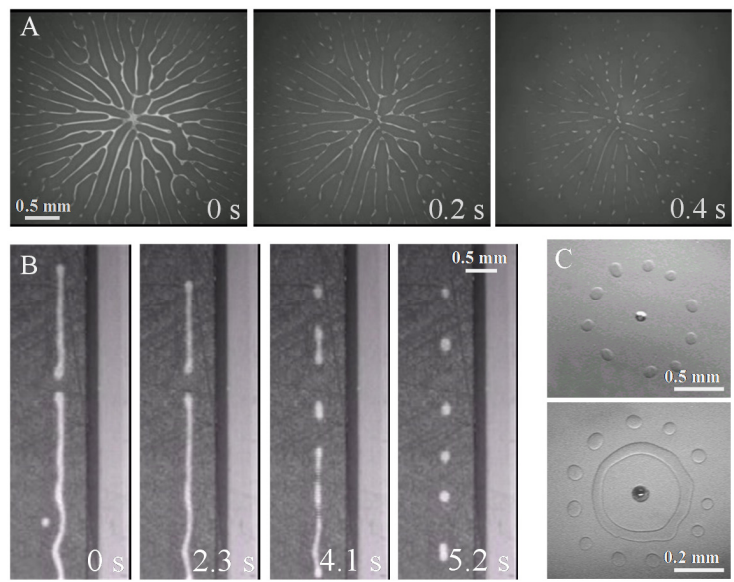

**Figure 15.** Various analogs of Rayleigh instability in thin elastic films A) Optical micrographs of the final stage of the healing of a blister [37]. B) A long air channel captured at the interface of a contactor and a thin film breaks up as bubbles. C) Concentric rings of air and bubble surrround [32] a tiny glass sphere (diameters 50-100 µm) as it separates a flexible cantilever and thin PDMS film. (C is reprinted with permission from Ref. [32]. Copyright (2003) The American Chemical Society).

At the final phase of the experiment, the cantilever does not come into contact smoothly with the PDMS film. But, a long air channel gets trapped at the same distance as where the bubbles were created in the first place. This long channel eventually undergoes an instability in the form of cavities. The sequence of events of this final state of the process is shown in figure 15B (Appendix B, Movie 3).

Experiments show that the spacing of the bubbles in all the above experiments is again about four times the thickness of the film, suggesting that the physics of this instability is also same as that discussed above. As the pressure inside the bubble increases, it enhances the flux of the air trapped from inside the channel outward. The bubbles, thus, can disappear sooner or later after their formation, unless it is counterbalanced by a tension in the plate. This is evident when the flexible plate is brought into contact with a thin soft confined film over a small particulate defect. The deflection ($w$) of the surface of the film can be described [105] by $\nabla^6 w \sim \left(E/DH^3\right)w$, which yields a damped oscillatory profile of $w$. The normal stress ($\sigma$) is also damped oscillatory (e.g. figure 12B) by virtue of the relation: $\sigma \sim \left(E/H^3\right)\nabla^{-2}w$. A solution of this equation yields a relationship between the moment applied on the crack tip [expressed as the applied force per unit length ($F$) times the crack length ($b$)] and the tensile stress generated slightly inside the edge of contact. When this stress reaches a critical value ($\sigma_c$), [i.e. $F_{\max}b = 3\sigma_c H\left(D/12\mu\right)^{1/3}$] cavitation [34] occurs. Surprisingly, this cavitation stress has been found to be only about 60 kPa for PDMS elastomeric films in contact with a silanized glass.

In cylindrical symmetry, the bubbles cavitate at the interface in the tensile zones as annular rings (figure 1). If the tensile stress is strong enough to counterbalance the driving force for closing the gap, the ring remains stable, which occurs usually in the first tensile zone of contact. The annulus degenerates into small hemispherical bubbles (i.e. it undergoes an equivalent Rayleigh instability) in the subsequent rings, as the tension decreases progressively.

We have thus far discussed several cases, in which an air bubble can be trapped at the interface of a thin flexible plate and a soft film that remains in an equilibrium state provided that there exists a force counteracting adhesion. From the equilibrium dimension of the bubble, an estimate of the work of adhesion is possible [37]. A similar method has indeed been exploited by Zong et al [106] to estimate the work of adhesion of a thin graphene film with silicon.

## 12. Meniscus Instability in Viscoelastic Films

While the relation $\lambda \sim H$ manifests in elastic films, a relation such as $\lambda \sim H\sqrt{\gamma/\eta V}$ governs the Saffman-Taylor instability [5] in a liquid, where $\gamma$ is the surface tension of the liquid, $\eta$ is its viscosity, and $V$ is the average displacement velocity of the liquid in a channel of height $H$. Nase et al [100] reported a beautiful study in which the transition from Saffman-Taylor to elastic instability was systematically observed and characterized. Here we present some additional intriguing observations regarding certain mixed mode instabilities as observed in experiments [97] with viscoelastic films. The films were made from incompletely cross-linked polydimethyl siloxanes (such as a commercial elastomer: Sylgard 184), the storage and loss moduli of which range from 1 to 3 kPa and 0.1 to 2 kPa, respectively. Figure 16 shows the patterns of instability which are obtained with such films in a cantilever geometry. Finger like patterns appear here as well; but it is evident from their fine structures that they are composed of long fingers with rounded tips that are accompanied with the small features that branch off normal to these long fingers.

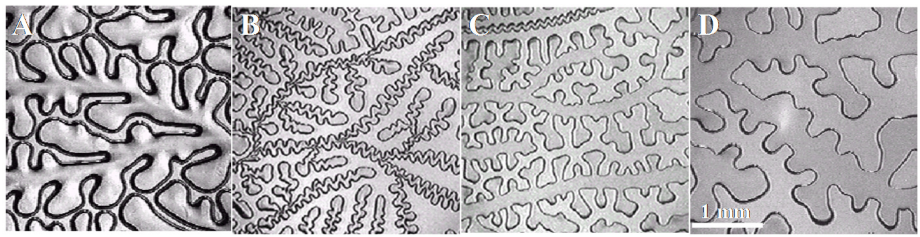

**Figure 16.** Experiments as in figure 12A were performed using viscoelastic PDMS films [97] bonded to a rigid substrate while a contacting plate of flexural rigidity, $D = 1.2$ Nm is lifted off the film quasi-statically. Optical micrograph (A) represents a film of $H = 38\,\mu$m prepared by using 1% by weight of curing liquid. Micrographs (B) – (D) represent partially crosslinked networks prepared by using 1.5% by weight of curing agent and of thickness, $H = 40$, 52 and $102\,\mu$m respectively. The separation distance between small side fingers as in images (B) – (D) follow the relation: $\lambda = 4H$ as with the purely elastic films.

These fingers do not appear instantly, but they evolve in the following progression. First a long finger appears, which does not remain stable, but it bifurcates to form small secondary side branches along its length. For very thin films, the long finger displays a telephone coil like structure as in micrograph 16(A). Eventually one of these side chains or coil elongates to form a new long finger that then branches out and evolves into more exotic patterns (figure 16). It is also intriguing that the long fingers appear somewhat randomly with no apparent characteristic separation distance. However, a long finger often suppresses the development of its neighbors thus resembling the viscous fingering instability, where such growth of one finger at the expense of the other occurs owing to a differential Laplace pressure. The short secondary fingers, nevertheless, remain uniformly separated, with the characteristic spacing increasing with thickness of the films as, $\lambda = 4H$. With these patterns, therefore, one can identify signatures of both viscous and elastic instability. The simultaneous appearance of (possibly) both viscous and elastic instability makes these patterns distinct from those observed by Nase et al [100] with somewhat thicker viscoelastic films, for which either Saffman-Taylor type viscous fingering or elastic instability patterns were observed without any transition between them. The thinness of our films makes them more susceptible to the effect of confinements which seems to evolve locally due to the nucleation and propagation of viscous fingers.

## 13. Instability in Stretched Film

There is a well-known surface instability called the Biot instability [13] that appears when a rubber block is compressed bi-axially or even uniaxially. This instability is a consequence of the competition between the in-plane and the out-of plane stretches on the surface of the rubber. At a critical compression, the decrease of the in-plane stretching energy overcomes the increase of the out-of plane stretching energy, thus causing the surface to undergo a morphological transition in the form of ripples or creases [56]. The critical compression ratio from variety of experiments is found to be about 0.3. This tendency of an instability developing from a uniaxial stretching can be taken advantage of in generating some non-trivial saw tooth like patterns in the asymmetric cantilever geometry. As discussed earlier, U shaped fingers develop when a flexible plate is peeled from a supercritically confined soft film. However, if the peel experiment is performed with a film thicker than a critical value, no instability develops. On the other hand if the same film is pre-stretched uniaxially, the increase of the out-of plane stretching energy can be further collaborated by a decrease of the elastic energy stored in the beam [97]. The tendency of the surface to develop crease like instability (but with the deflection of the surface protruding outward) with its wave vector perpendicular to the direction of the initial stretch of the film increases the amplitude of the finger like instability. The net result is the development of a saw tooth like instability at the crack front as shown in figure 17.

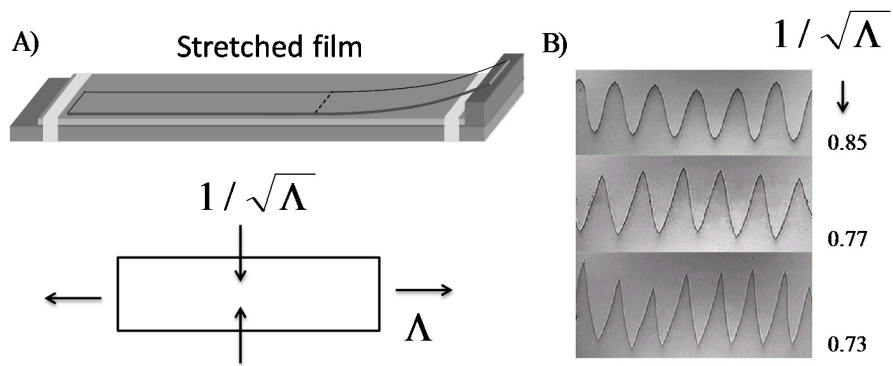

**Figure 17.** Patterns of instability for a pre-stretched film. (A) Schematic shows an elastic film, first stretched to a desired extension ratio $\Lambda$ and then bonded to a rigid substrate. B) "V" shaped fingers are observed [99] along the contact line when a flexible plate is placed over the stretched film. . (Reprinted with permission from Ref. [99]. Copyright (2007) Taylor & Francis Inc.)

There is however a difference between the saw toothed instability described here and the creasing instability. Although the creasing instability occurs at a critical compressive ratio of ~ 0.3 that being independent of the shear modulus of the material or its dimension, the critical compressive strain for the saw tooth instability to occur in the stretched film as induced by adhesion varies with both these parameters in addition to the flexural rigidity of the contacting plate. For example, with an elastic layer with shear modulus $\mu = 1.0$ MPa and thickness $H = 645$ μm, flexural rigidity of plate, $D = 0.02$Nm, the instability appears when the compression ratio $\Lambda_2 = 1/\sqrt{\Lambda_1}$ diminishes below 0.85, which is much larger than the critical value of the creasing instability.

## 14. Bubble Motion due to Strain Energy Gradient

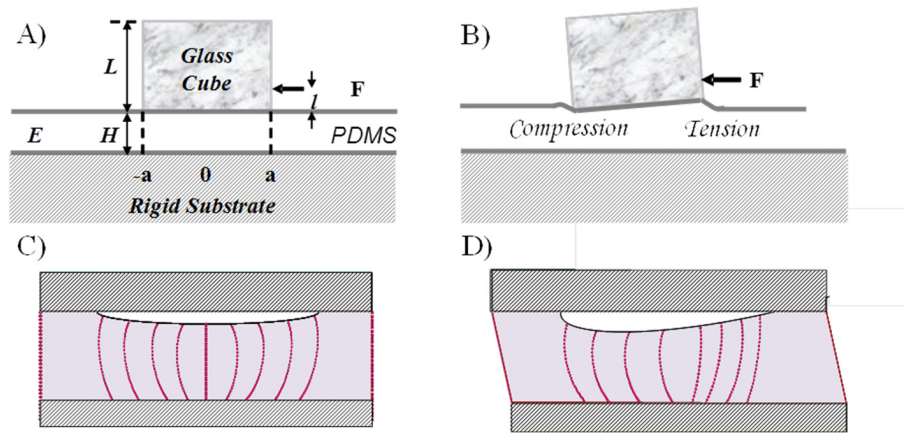

**Figure 18.** (A) and (B) Schematic of the shear experiment of a silanized glass cube against a thin elastic film bonded to a rigid substrate. Since the force is applied parallel to the interface, but slightly above it, a torque is developed that creates a tension at the rear edge, but a compression near the advancing edge. (C) and (D) Schematic of the film at the vicinity of a cavity. The solid lines represent the horizontal displacement of the film with respect to that (zero) of the lower interface. This symmetric pattern becomes asymmetric when a shear force is applied that causes crack to heal at one end but it opens up at the other end.

One intriguing feature of the cavities formed via elastic instability is that they could be quite mobile at the interface when the contactor is slid against the thin film [107]. These movements of the cavities are the thin film counterparts of the well-known Schallamach waves [108-111], which develop when a rigid indenter slides against a thick rubber slab. In our experiments, the bubble like instability that develops at the interface is comparable in size to the thickness of the film. They always originate at the trailing edge of the slider-film interface, but they moves towards the advancing edge. The strain field around a bubble is symmetric when no tangential force is applied upon the film. However, when the upper block is translated against the film, the superimposition of the shear deformation of the film and that surrounding the bubble creates an asymmetric field that causes the crack to heal near the trailing edge, but to open up near the leading edge (Figure 18). This asymmetry causes the bubble to move from the trailing to the leading edge of the sliding interface (Appendix B, Movies 4,5).

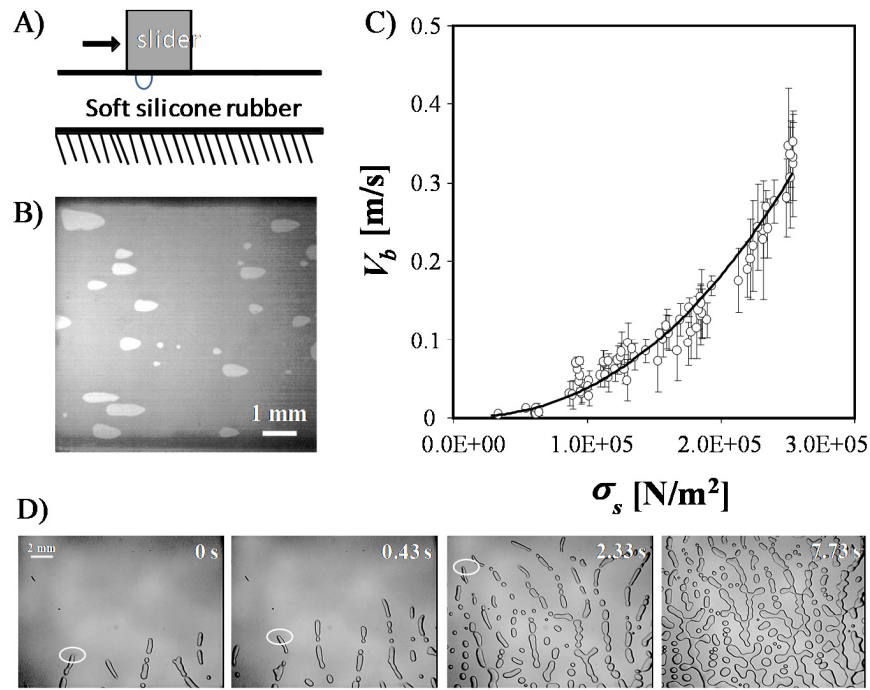

**Figure 19.** A) Schematic of a cavity produced at one end, moves towards the other end. (B) Video micrograph of the motion of the bubbles at the interface of a rigid disc and a soft PDMS film. C) The velocity ($V_b$) of the bubble varies [39] quadratically with the applied shear stress ($\sigma_s$). D) Image sequences of the formation of an interfacial pattern between a thin film of soft gel ($\mu\sim 10$ Pa) and a rigid glass plate that also shows the motion of bubbles. The white circles show how a stream of bubbles traverses across that coalesce and form a more complex pattern. This experiment was performed in a wedge shaped geometry (around the region of thickness~120 μm) described in Figure 8 (Appendix B, Movie 6). . (B & C are reprinted with permission from Ref. [39]. Copyright (2007) The European Physical Journal)

The main driving force for the motion of the bubble comes from the shear strain energy release rate $\sigma_s^2 \delta / \mu$, which is opposed by a viscoelastic drag in the rubber (here, $\sigma_s$ is the shear stress and $\delta$ is the height of the bubble that is usually much smaller than its diameter) and that arising from the dissipative processes occurring near the crack tip region. The latter is of the form $\sigma_0^2 V_b T / \mu$ (here, $\sigma_o$ is the cohesive stress near the crack, $V_b$ is the bubble velocity and $T$ is the relaxation time of the polymer) according to the cohesive zone model of Greenwood [112]. However, as the bubble moves, the rubber itself undergoes an oscillatory motion due to its deformation and the subsequent relaxation that creates a drag force of the form: $\mu \delta (V_b / V_0)$ ,where $V_o$ is a material velocity. By combining the above two factors, the bubble velocity can be expressed as:

$$V_b \sim V_0 \frac{\sigma_s^2}{\left(\sigma_0^2 V_0 T / \delta + \mu^2\right)} \qquad (26)$$

Experiments show that $V_b$ is nearly independent of the film thickness, but it increases linearly with $\sigma_s^2$, thus suggesting a couple of scenarios (Figure 19). For example, it may be the case that the bubble velocity is controlled mainly by the energy dissipation in the bulk of the rubber, or that the term $\sigma_0^2 V_0$ is on the order of $\mu^2 \delta / T$ .The latter would be the case if the cohesive stress $\sigma_0$ is on the order of the shear modulus $\mu$ and that $V_0$ is on the order of $\delta / T$. All these considerations lead to an expression of the bubble velocity as:

$$V_b \sim V_0 \left(\frac{\sigma_s}{\mu}\right)^2 \qquad (27)$$

Experiments show that $V_b \sim (\sigma_s / \mu)^2$ with $V_0$ having the value of 10 m/s, which is one third of the shear wave velocity of the rubber. Thus, if the applied shear stress is on the order of the shear modulus of the rubber, the bubble velocity would approach its shear wave speed, which would, of course, be tempered by the inertial force as is the case with the high speed fracture. Since the shear stress at the rubber substrate interface is mainly kinetic, i.e. it grows sub-linearly with the sliding velocity as $\sigma_s \sim V_s^{0.5}$, we expect that the bubble velocity to be proportional to the sliding velocity. Experimentally, we find this ratio of $V_b$ and $V_s$ to be on the order of thousands, which is much larger than what is typically observed with the propagation of the Schallamach waves. For example, Rand and Crosby [110] found this ratio to be around 8, whereas Viswanathan et al [111] found it to be around 30 to 50. These differences are, of course, related to the differences in the viscoelastic properties of the rubbers used in the two types experiments.

## 15. Summary, Perspective and Suggestions for Future Studies

In this section, we re-iterate the main points of this article and highlight certain issues that are still unresolved, or poorly resolved, but are worthy of consideration to make further progress in the field. We also take this opportunity to provide some guidelines with which some new instability driven research could be initiated for future studies.

The main point of this article is to illustrate how a soft confined film relieves the imposed constraint by undergoing an elastic instability that gives rise to various types of morphological patterns. In this de-bonding mode, the decrease of the potential energy of the system creates the main driving force for instability and pattern formation in thin soft films. The origin of the force giving rise to the decrease of the potential energy is somewhat immaterial. Indeed the relationship $\lambda \cong 3H$ has also been found recently in the gravity induced instability of a hydrogel layer [55] of finite thickness.

 The detailed fine structure of the pattern depends on the distribution of stress that leads to various elastic counterparts of the generic instabilities observed with viscous liquids. We also discussed how these instabilities serve such useful functions as controlling transport via self-generated channels, allowing sliding by auto-roughening the interface and facilitating debonding of a rigid contactor by enhancing the overall compliance of the joint. The scaling laws for the wavelength of elastic instability can be understood in terms of the film thickness, co-operativity, and the relative contribution of the elastic modulus to the surface tension of the film. When the surface tension predominates over elasticity, the instability exhibits a much longer wave feature, which plays a crucial role in the debonding of a very soft or a very thin adhesive joint that warrants a modification of the scaling laws for fracture from that of the well-known Griffith theory [69].

These studies also have important implications in understanding the delamination of viscoelastic adhesives from a solid substrate [113], in which the wavelength of the fingering pattern is rather insensitive to the peel speed. In order to elucidate this particular feature, Fields and Ashby [92] had to invoke a highly non-linear constitutive law of the viscous deformation of the adhesive. The studies performed by us as well as by Nase [100] et al now strongly indicate that many of these instabilities are, more or less, elastic in origin. The fact that the amplitude of instability in a peel configuration depends on various material properties calls for an in-depth non-linear stability analysis of the problem that would allow a more comprehensive understanding of the relationship between adhesion and the fingering amplitude. When such an analysis is in place, the amplitude of instability could indeed be used as a reporter to the strength of an adhesive interface, which had, already, been evident in a previous publication [51]. These studies are expected to shed more light onto other types of instabilities known for viscoelastic systems as well, especially those displaying shark-skin [96] or saw-tooth patterns [48].

At this juncture, we present an amusing, albeit incomplete, similarity of the wavelength of elastocapillary instability in the cantilever geometry [29] as discussed in the main text with an approximate expression of the same using the method proposed by Babchin [94] et al and Brown [95]. According to these authors the wavelength of instability is: $\lambda \sim \sqrt{(\gamma / \sigma_x)}$, where $\sigma_x$ is the gradient of the applied stress. In the peel geometry involving a soft elastomeric film, we may write: $\sigma_x \sim \mu w_0 / \beta H$, $w_0$ being the normal displacement of the elastomer at the crack tip and $\beta$ is the length over which the deformation stress decays along the interface of contact. As shown in reference 29, $w_0$ is related to the work of adhesion ($W_a$), the modulus of the film ($\mu$) and its thickness as: $w_0 \sim (\beta/H)(W_a H/\mu)^{1/2}$. Using the above expressions for $\sigma_x$ and $w_0$, we have $\lambda \sim (\gamma W_a)^{1/4}(\gamma H^3/\mu)^{1/4}$, which, for $\gamma \approx W_a$, becomes same as that obtained from a rigorous analysis discussed in section 6.

While all the analyses of instability, so far, has been performed for a purely elastomeric film or a gel without any energy dissipation, a question, nevertheless, arises as to what extent the viscous dissipation in the film influences the wavelength selection process. This is an important question to address as a vast majority of adhesives are dissipative. Here, we provide a rudimentary outline of how such a problem can be tackled by balancing the rate of the work done by the external load with the rate of dissipation of the energy due to a viscous process as follows:

$$\left(-\frac{\partial U}{\partial w_0}\right)\dot{w}_0 \sim \eta H L \int \left(\frac{\partial \dot{u}}{\partial z} + \frac{\partial \dot{w}}{\partial x}\right)^2 dx \tag{28}$$

Where, $\eta$ is the viscosity, the dots above $u$ and $w$ indicate time derivative, and $U$ is:

$$U \sim -Fw_0 + \mu H L \int \left(\frac{\partial u}{\partial z} + \frac{\partial w}{\partial x}\right)^2 dx + \gamma L \int \left(\frac{\partial w}{\partial x}\right)^2 dx \tag{29}$$

Equations 28 and 29 admit a solution for $w_0$ as follows:

$$w_0 = \frac{\sigma\left[1 - \exp\left\{-\dfrac{2\mu H f_1(\lambda, H) + 2\gamma f_2(\lambda)}{\eta H f_1(\lambda, H)}t\right\}\right]}{2\mu H f_1(\lambda, H) + 2\gamma f_2(\lambda)} \tag{30}$$

With, $f_1(\lambda, H) = \left(\lambda/H^2 + 1/\lambda\right)^2$ and $f_2(\lambda) = \left(1/\lambda\right)^2$. The maximum growth rate of $w_o$ at short time ($t \sim 0$) occurs at $\lambda \sim H$, which results mainly from the minimization of viscous dissipation. However, as time progresses, especially at equilibrium ($t = \infty$), the maximum value of $w_0$ occurs at $\lambda \sim H\left(1 + \gamma/\mu H\right)^{1/4}$ that is precisely same as what has been obtained previously resulting from the minimization of elasto-capillary energy. A crossover of the two types of behaviors is expected to occur at some specific time, which would require a detailed formal analysis of the problem. There are also other possibilities to consider. For example, it is possible for the cavities to channel laterally creating a hydrostatic pressure gradient that, in conjunction with the surface tension forces, can give rise to a Saffman-Taylor like scenario thus resulting in the coexistence of two scales of instabilities. A cogent analysis of the problem delineating different manifestation of the two types of instabilities has been carried out by Nase et al. [100]

At this juncture, let us point out that both the scaling relations $\sigma_c \sim \left(W_a E / H\right)^{1/2}$ and $\sigma_c \sim \left(\gamma / EH\right)^{0.25} \left(W_a E / H\right)^{0.5}$ can also be derived from equation 5 by replacing $a$ with the wavelength of instability, which are $\lambda \sim H$ for a high modulus elastomer and $\lambda \sim \left(\gamma H^3 / \mu\right)^{1/4}$ for a low modulus gel respectively. We may thus expect that the general scaling relation for the debonding stress can be obtained by using the interpolating expression for the wavelength, thus yielding $\sigma_c \sim \left(1 + \gamma / EH\right)^{0.25} \left(W_a E / H\right)^{0.5}$.

The above studies prompt a natural question: what role can surface tension play in a semi-infinite film, where there is no characteristic length scale of instability. We discuss this point below with a scenario that may also have practical implications. We consider a rigid stud composed of parallel cylindrical fibrils being pulled off from a semi-infinite half space of a soft polymer (figure 20).

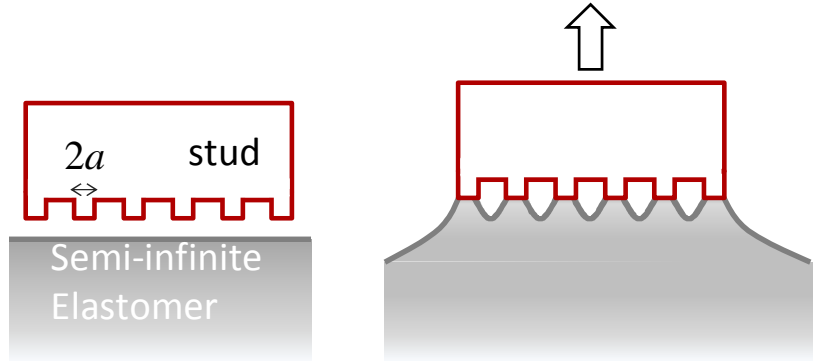

**Figure 20**. Schematic of the interfacial morphology when a fibrillated rigid stud is pulled off a semi-infinite elastomer.

For simplicity, let us consider that the radius of each of the fibril and their spacing is $a$. In this problem if the elasto-capillary number ($\gamma/Ea$) is not negligible, surface tension would contribute to the overall stiffness of the adhesive as the surface would corrugate during the pull-off process. At a scaling level, starting with the total energy ($U$) of the system (composed of potential, elastic [79] and surface energies, equation 31), the unstable crack growth condition

$$U \sim -F^2/(Ea + \gamma) + W_a a^2 \tag{31}$$

predicts that the critical pull-off stress is: $\sigma_c \sim (1 + \gamma/Ea)(W_a E/a)^{0.5}$. Since the elastocapillary number $\gamma/Ea$ can be enhanced by decreasing the radii of the fibrils, significant control of the fracture toughness may be obtained in the pull-off process.

On the issue of the debonding stress, one notices that for an ultra-soft gel [77] a plot of $\sigma_c (EH/\gamma)^{0.25}$ versus $(W_a E/H)^{0.5}$ allows the experimental data to cluster around a master line, although the slope ($\sim 26$) of the resultant line is much larger than that ($\sim 1.9$) obtained from the plot of $\sigma$ versus $(W_a E/H)^{0.5}$ for the higher modulus elastomers. One obvious consideration behind the high stress observed with a hydrogel gel is that there may be an underlying viscoelastic amplification of the Griffith's stress that has not been taken into account. Such an amplification of fracture stress has recently been observed by Nase et al [114] with partially cross-linked PDMS films. However, as the hydrogel films, in our experiments, are very thin, it may not allow sufficient dissipation of energy for the viscoelastic amplification to be effective. In fact, by varying the pull-off speed over a factor of thousand led to an increase of the pull-off stress by only a factor of 2 to 3 with the soft hydrogel film, which is insufficient to explain the observed discrepancy.

These observations suggest that we need to go few steps beyond the simple energy balance approach used here and study closely the structure of the three phase contact line for the ultra-soft gels. Previously, it was suggested [51] that the contact angle near the crack tip for a soft adhesive could be controlled by the modulus of the polymer and the frictional stress ($\sigma_s$) near the contact line region as: $\tan\theta = 2\mu / \left[ \sigma_s (1 - (\mu / \sigma_s)^2 \right]$. If the frictional stress is on the order of the modulus of the material, the three phase contact angle can easily approach $\pi / 2$ thus thwarting the stress intensity at the crack tip further, thereby increasing the elastic energy dissipation by allowing additional extension of the gel. Fracture, however, is fundamentally a non-equilibrium process that may not be described by the instability modes obtained from an energy minimization process. Indeed, using a toy model, in which an uniform distribution of certain numbers of modes are specified to participate in the fracture simultaneously, it becomes quite easy to explain the high value of the pull-off stress observed with the soft gels. Here, the role of viscosity may just be to delay the decay of the non-principal modes. A more realistic situation would, however, be to develop a criterion that would impart an appropriate weightage to a specific mode, from which a reliable estimate of fracture stress can be obtained.

Another issue that requires amicable resolution is the stress at which cavitation occurs at the interface of the film and the contactor. Experiments of the types described in figure 12 lead to a value of $\sigma_c \sim 60$ kPa at the interface of PDMS and a silanized (low energy) glass, which is indeed low. It will be easy to understand this low stress if this apparent cavitation corresponds to the expansion of internal cracks from defects as discussed by Shull and Creton [45]. We, however, point out that the surfaces used in out experiments are rather smooth with a root mean square roughness of the contacting surfaces ~3 Å. The result is also independent of whether these studies are performed with a pre-crosslinked elastomer debonding from a rigid substrate or the debonding is performed with an elastomer after it is crosslinked against the rigid substrate from liquid state. Furthermore, cavities do not appear in random manner, but they appear with a well-defined spacing. Therefore, it is not very clear in our experiments if the defects at the interfaces give rise to a simple propagation of an existing crack or that some kind of nanoscale interfacial instability (mechanism of which is not yet known) lead to stress concentration leading to initiation of the cavities. This could be an interesting topic for future investigations.

We anticipate that similar line of investigation with hydrogel and other soft material interface could provide a more succinct picture to the problem at hand and other related vexing issues. If the cavitation stress could be so low, one may expect that the cohesive stress at the soft matter interfaces may also be much lower than that predicted by the van der Waals stress, which is, indeed, consistent with couple of recent studies [99,115,116]. By developing such an argument even further, there may exist an opportunity to settle certain unresolved issues of the fracture involving soft material interfaces. In this context, a low modulus rubber and a hydrogel present a contrasting scenario as the latter may be accompanied by the cavitation of water at the interface, which is known to occur at Mega-Pascal range of hydrostatic tension in the bulk (as pointed out by one of the referees).

An issue of sufficiently fundamental consideration related to the fracture with certain neo-Hookean materials is that a normal stress difference ($\tau_{11} - \tau_{22}$) may develop near the crack tip region. A product of this quantity and an effective thickness ($\delta$) of the active zone near the crack tip yields a local strain energy release rate. What is the exact role of this normal stress difference in the fracture of soft matter interface has still not been worked out, even though the idea was proposed [117] quarter of a century ago.

Although an investigation of the static properties of the morphological evolution of patterns provides an avenue to study the nucleation of a crack, focusing on the dynamic properties at soft matter interfaces could also shed more light on the propagation of interfacial dislocations, and thus dynamic friction. Furthermore, interfacial dynamics is the enabling science underlying the design of soft robotic systems, where friction is an important issue. By combining the role of confinement with the low dynamics friction of a judiciously selected elastomer, it is indeed possible to firmly attach a flat object to a thin soft film that can be slid or rotated [39] afterwards without any additional normal force acting on the object. Cavitated bubbles do, however, appear at the interface, which sometimes give rise to stick-slip instability. Although, the instability can be eliminated by creating pre-existing air channels on the surface of the film, studying the dynamics of these cavities provides a means to probe interfacial dynamics. We expect that an in-depth insight into the propagations of interfacial dislocations [118-124] in soft matter systems can be gained by augmenting the scope of these studies with other complementary experiments. In this context, we propose two additional lines of investigations as discussed below.

### 15.1. An Outline of an Experiment to Study Interfacial Waves in Supported Thin Film

A first line of investigation can be carried by sliding a hemisphere laterally against a thin elastomeric film (~ 5-10 $\mu$m) supported on an incompressible liquid thus forming wrinkles that could pass through the area of contact as pulses (Figure 21 & Appendix B, Movie 7). Since the spacing of the wave pulses is governed by a long wavelength instability, the dynamics of a single wave pulse can be studied as a function of the surface property of the slider, the thickness of the film and its viscoelastic property. This moving wrinkle could also be a nice model system to study the canonical problem of the motion of an elastic line through the interface that may be pinned via double or multivalued energy wells.

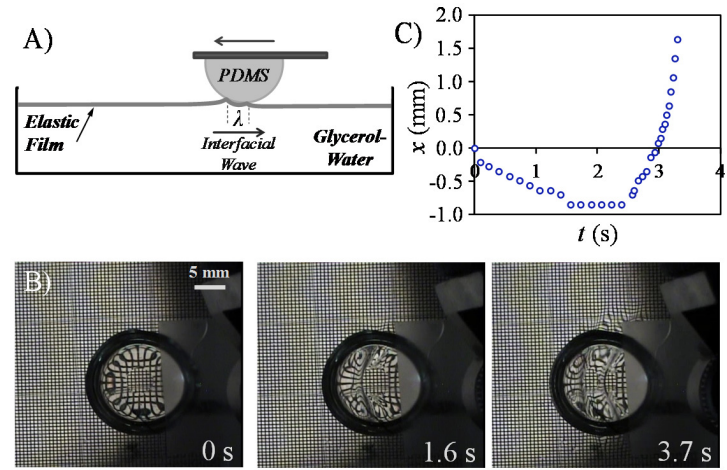

**Figure 21.** (A) Interfacial waves at the interface of a rigid hemisphere and a thin (7.5 μm) films of silicone elastomer supported on the surface of an incompressible liquid (1:1 mixture of water and glycerin). When the spherical indenter is slid past the elastomeric film of thickness, a single wrinkle develops, which propagates through the interface. The sequences of the motion of the wrinkle are shown in B. The lateral displacement of the wrinkle as a function of time is shown in figure C. The spacing of the wrinkle pulses is governed by the long wavelength: $\lambda = 2\pi(D/\rho g)^{1/4} \sim 1.6$ mm, $D$ being the bending modulus of film, $\rho$ the density of the liquid and $g$ the acceleration due to gravity.

## 15.2. Can Instability be Used to Generate Motion?

In 1981, Kendall [125] published an intriguing paper, in which he studied the formation of dislocation of a flexible substrate in the form of a fold by using a lateral constraint and implicated its role in the toughening of adhesive composites. A subsequent, and a closely related study [126], exemplified how the sequential detachment and attachment of interfacial waves enhance the fracture energy of a rubber strip in a low angle peeling. More recent studies have explored the dynamics [127] of the motion of a deliberately formed fold in the rubber strip that moves due to the action of gravity. Folds in a rubber strip can also be formed by taking advantage of a kink instability, which is an energetic process [128,129] that even generates surface waves [128]. Based on these prior studies, here we present an outline of an experiment, in which a folding instability (figure 22) can be evoked in a thin polymer strip supported on a thick rubber. At the onset of the kink instability [128,129] in the thick rubber, part of the strain energy is shared by the thinner strip, which subsequently detaches from the rubber slab in a snap out mode. The resultant fold moves unidirectionally if the strip is endowed with a gradient of thickness (thus, a gradient of the strain energy density) (Appendix B, Movie 8). An intriguing related phenomenon is that alternate bending and relaxation of the rubber block often causes the entire strip to translate on the rubber slab, which may be thought of as an elastic analog of the surface energy gradient driven motion of drop [130] on solid surfaces.

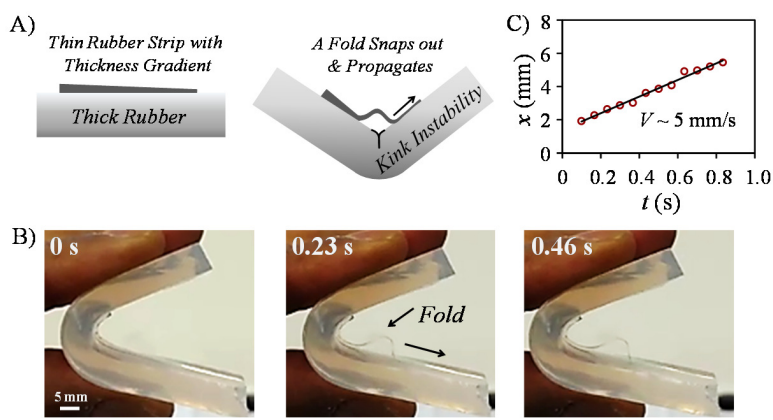

**Figure 22**. (A) Schematic of the formation a fold in a thin film as triggered by a kink instability in a thicker rubber slab of silicone elastomer (70mm x 20mm x 8mm). A thinner strip of silicone elastomer (28 mm x 2 mm with a thickness gradient of 0.5 mm/1 mm) is placed above the thick slab. (B) A kink develops in the thicker slab when it is bent, that leads to a snap out instability in the thin rubber strip. Because of the gradient of a strain energy, the fold moves from the thicker to the thinner part of the rubber. Repetition of this process often causes the entire strip to translate on the thicker rubber slab. (C) Displacement of the closing edge of the fold is plotted as a function of time. (These experiments were performed with Saheli Biswas).

### 15.3. Can Instability be used to Arrange Particles?

We end this article by pointing out an interesting possibility that could emerge with multiple particles sprinkled on the free surface of a very thin film, in which its free surface as well as the particles could interact with the underlying film/substrate interface via long range van der Waals forces (Figure 23).

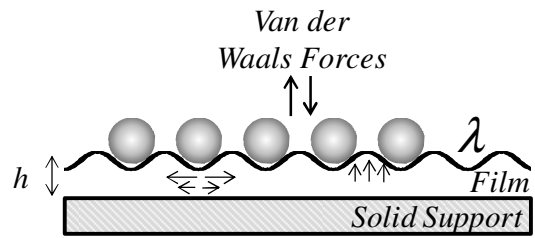

**Figure 23**. Schematic of the organization of particles due to the combined effects of elasticity and van der Waals forces.

Considering that the thickness of the film undergoing a periodic perturbation: $h = h_o - w_o \cos(k_2 x)$, the excess van der Waals free energy can be written as follows:

$$\Delta G = \frac{1}{L_2} \int\limits_0^{L_2} \left[ \frac{dG(h,x)}{dh} \Delta h + \frac{1}{2} \frac{d^2 G(h,x)}{dh^2} (\Delta h)^2 \ldots \ldots \right] dx \qquad (32)$$

With a periodically varying Hamaker constant, $\mathcal{A}(x) = \mathcal{A}_o \cos(k_1 x)$, $\Delta G$ is:

$$\Delta G = -\frac{\mathcal{A}_o w_o}{6 \pi L_2 h_o^3} \int\limits_0^{L_2} \cos(k_1 x)\cos(k_2 x)dx - \frac{\mathcal{A}_o w_o^2}{4 \pi L_2 h_o^4} \int\limits_o^{L_2} \cos(k_1 x)\cos^2(k_2 x)dx \qquad (33)$$

When $k_1 = 0$, $\Delta G$ is controlled by the perturbation term of $\Delta G$. On the other hand, with $k_1 = k_2$, the first term on the RHS of equation 29 governs the interaction. Under any of these situations, the deposited particles are expected to roll and/or slide on the surface (provided that the friction is negligible) to adjust their positions so as to minimize the elastic energy at the gain of the energy of interaction. Even more intriguing possibilities could manifest with a thin film of a liquid crystalline film [131,132], where the balance could be between the van der Waals force and the elastic force brought about by the alteration of the director field in the liquid crystal by the particles themselves.

**Acknowledgements:** MKC is forever grateful to (late) Professor Alan Gent for encouraging him to carry out an in-depth study of elastic instability in thin films. Prof. Gent originally suggested to MKC (ca. 1994) that an analog of Saffman-Taylor instability would be found in pure elastic films. That suggestion is the kernel of all that is reported here. We thank Prof. Yves Pomeau for bringing reference 2 to our attention. We also benefited from numerous discussions with several of our colleagues over the years. We are grateful to the Office of Naval Research and Dow Corning Corporation for supporting this work.

**Author contributions:**

All authors contributed equally to the paper.

## Appendix A: List of Symbols

$\lambda$ (m): Wavelength of instability

$H$ (m): Thickness of thin elastic/viscoelastic films

$\mu$ (Pa): Shear modulus

$E$ (Pa): Elastic Young's modulus

$\mathcal{A}$ (J): Hamaker's constant

$\ell$ (Å): the molecular separation distance between the two surfaces

$\sigma$ (Pa): Normal stress Stress

$\sigma_c$ (Pa): Critical debonding stress

$u, v, w$ (m): Deflections of the surface in the $x, y, z$ directions

$v$ : Poisson's ratio

$\nabla$ : Differential operator, which is $\nabla \equiv \partial/\partial x$ in one dimension

$r$ (m): Radial distance

$a$ (m): Radius of the cylindrical stud

$U_T, U_P, U_E, U_a, U_S$ (J): Total, Potential, Elastic, Adhesion, Surface Energies respectively

$\Pi_T, \Pi_{elastic\,(film)}, \Pi_{bending}, \Pi_{adhesion}$ (J): Energies

$F$ (N): Applied force/load

$W_a$ (J/m$^2$): Work of adhesion

$\gamma$ (N/m): Surface tension

$A$, $A_{finger}$ (m$^2$): Interfacial Area of contact

$k$ ($\equiv 2\pi/\lambda$) (m$^{-1}$): Wavenumber

$M$ (N/m$^3$): Stiffness of the elastic film, which is a function of the wave number $k$

$A_i, B_i$ : Coefficients in the Kerr's equation that are functions of the Poisson's ratio

$D$ (N.m): Bending rigidity

$\beta$ (m): Characteristic stress decay length

$\varepsilon$ : Control parameter ($\varepsilon \equiv H/\beta$)

$A_m$ (m): Amplitude of instability

$\xi$ : Non-dimensionalized amplitude ($\xi \equiv A_m/\beta$)

2c (m): Width of the stressed zone

$k_D$ (m$^2$): Darcy's permeability

$b$ (m): Crack length

$\psi$ (m): the vertical displacement of the plate from undeformed surface of film

$\eta$ (Pa.s) : Viscosity

$\Lambda$ : stretch extension ratio

$\delta$ (m): the height of a bubble at an elastic interface

$V_b$ (m/s): Bubble velocity

$V_0$ (m/s): Material velocity

$T$ (s): Relaxation time of the polymer

$\sigma_o$ (Pa): Cohesive stress near the crack

$\sigma_s$ (Pa): Shear stress

$\sigma_x$ (N/m$^3$): Gradient of applied stress

$\theta$ (rad): contact angle near the crack tip for a soft adhesive

$\Delta G$ (J) : Excess free energy

## Appendix B: Movie Captions

Movie 1: Formation of worm-like instability patterns at the interface of an elastic film and a rigid disc.

Movie 2: An air blister trapped between a thin elastic film (50 μm) and a glass cover slip slowly heals with time as the air escapes through the thin air channels.

Movie 3: Closing of an interfacial crack between an elastic film (195 μm) and a glass substrate, as the height between the upper glass substrate and the elastic film is gradually decreased. A thin air channel, towards the end, becomes unstable splitting itself to form several bubbles of somewhat uniform size, demonstrating a Rayleigh-like instability.

Movie 4: Interfacial bubbles moving across the interface of a smooth elastic film (127 μm) and a glass (slider). Sliding speed of the slider is 150 μm/s.

Movie 5: Interfacial bubbles moving across the interface of a structured elastic film (127 μm) and a glass (slider). Sliding speed of the slider is 150 μm/s.

Movie 6: Motion of cavitated bubbles at the interface of a glass slide and a soft hydrogel in a peel configuration (for details see figure 19).

Movie 7: Propagation of wrinkling instabilities in a thin elastic film (7.5 μm) supported over aqueous glycerol due to the sliding of a hemispherical PDMS slider (diameter 18 mm).

Movie 8: Kink instability in a thick elastomeric slab induces the formation of a fold in a thin elastomeric film (having a thickness gradient) and the folding instability propagates due to strain energy gradient.